\documentclass[
  reprint,
  amsmath,
  amssymb,
  aps,
  prx,
  showkeys,
  floatfix
]{revtex4-2}

\usepackage[version=3]{mhchem} 
\usepackage{textgreek}
\usepackage{graphicx}
\usepackage{siunitx}
\usepackage{indentfirst}

\newcommand{\abbreviations}[1]{\gdef\storedabbreviations{#1}}
\newcommand{\printabbreviations}{%
  \begingroup
  \centering
  \textbf{Abbreviations:} \storedabbreviations\par
  \endgroup
}

\newenvironment{suppinfo}{%
  \clearpage
  \onecolumngrid
  \section*{Supplemental Material}%
}{%
  \clearpage
  \twocolumngrid
}

\begin{document}

\title[Lipid-Mediated Shape Changes in Liquid Crystal Droplets]
  {Lipid-Mediated Control of Thermally Induced Shape Transformations in Liquid Crystal Droplets
}

\author{Mengwei Li}
\affiliation{Soft Condensed Matter and Biophysics, Debye Institute for Nanomaterials Science, Utrecht University, 3584 CC Utrecht, The Netherlands}

\author{Lisa Tran}
\email{l.tran@uu.nl}
\affiliation{Soft Condensed Matter and Biophysics, Debye Institute for Nanomaterials Science, Utrecht University, 3584 CC Utrecht, The Netherlands}

\abbreviations{8CB,5CB,DLPG,DOPG,POM}

\begin{abstract}

\noindent Understanding how molecular-scale organization is translated into mesoscale shape transformations remains a central challenge in soft matter. Lipids and liquid crystals provide a useful platform for addressing this question because variations in hydrocarbon-chain packing can modify interfacial order, anchoring conditions, and elastic stresses, potentially coupling molecular organization to droplet morphology. Here, we investigate the temperature-dependent structural evolution of monoolein-doped 4-octyl-4'-cyanobiphenyl (8CB) droplets dispersed in aqueous phospholipid solutions. Using polarized optical microscopy, we show that the internal monoolein concentration and the external lipid environment jointly regulate phase transitions, thin filamentation, and larger deformations during heating. The droplets undergo coupled smectic-nematic-isotropic transitions, with extended thin filaments observed exclusively in the smectic regime. At the smectic-to-nematic transition, we observe an abrupt and discontinuous onset of droplet shape deformation, revealing a shape-change transition coupled to the bulk mesophase transition. To our knowledge, this is the first report of a discontinuous droplet-shape-change transition coincident with a smectic-to-nematic phase transition. Varying the internal and external lipid contents redirects the reconfiguration pathway between filament-dominated smectic responses and more amorphous nematic shape changes. Interfacial tension measurements further show that stronger reductions in liquid-crystal-aqueous interfacial tension do not necessarily produce filamentation or deformation. Instead, the observed morphodynamics arise from the coupling between bulk elasticity, mesophase structure, and lipid-mediated interfacial organization. These findings establish lipid composition and hydrocarbon-chain architecture as key parameters governing thermally induced shape transformations in liquid crystal droplets.
\end{abstract}

\maketitle
\printabbreviations

\section{Introduction}

Droplet shape transformations provide a route to connect molecular organization with emergent mesoscale structure. In simple isotropic systems, interfacial tension drives droplets toward spherical shapes by minimizing surface area. Under suitable conditions, however, droplets can develop anisotropic morphologies such as filaments, protrusions, facets, or branched structures \cite{Eggers2008,Squires2005,Denkov2015,Guttman2016,Liber2020,Haas2017,Marin2020,GarciaAguilar2021}. These studies show that droplet morphology can be strongly altered when interfacial forces are coupled to phase transitions and elastic stresses. Liquid crystal droplets are particularly useful in this context because interfacial forces, orientational elasticity, topological constraints, and thermally accessible phase transitions have comparable energy scales \cite{deGennesProst1995,LopezLeon2011,Poulin1997}. As a result, liquid crystal droplets can convert subtle molecular perturbations into pronounced morphological responses, making them valuable model systems for adaptive soft materials. Their ability to undergo controlled structural reconfiguration has also motivated applications in self-assembly and responsive optical systems \cite{Wang2016,Peddireddy2013OptExpress,Kim2019,Neogi2026}.

A major unresolved question concerns how the molecular structure of amphiphiles, particularly lipid hydrocarbon chains, influences droplet shape evolution. In lipid assemblies, hydrocarbon-chain conformation affects packing density, interfacial order, membrane elasticity, and spontaneous curvature \cite{Seifert1997,Sakashita2012,Ziherl2005}. Chain length, saturation, and conformational freedom are also well known to regulate membrane morphology and shape transitions in biological and biomimetic systems \cite{Hamley2007,Buddingh2017,Seifert1997,Sakashita2012,Ziherl2005}. Less is known, however, about how lipid-chain architecture controls the coupling between interfacial organization and adjacent elastic materials. This coupling is important because molecular packing at an interface can modify elastic stresses and bulk structures, providing a route through which molecular-scale changes are amplified into mesoscale morphological responses. Thermotropic liquid crystals provide a useful model system for investigating this problem because their phase transitions involve substantial changes in molecular packing and elasticity. In particular, the smectic-to-nematic transition removes translational order while retaining orientational order, creating an opportunity to examine how lipid-mediated interfacial organization couples to bulk ordering and shape transformation.

Recent studies have demonstrated that amphiphile adsorption can reshape liquid crystal droplets by modifying anchoring conditions, interfacial free energy, and elastic stresses. Bahr showed that surfactant adsorption can induce a nematic wetting layer at thermotropic liquid crystal-water interfaces, highlighting the sensitivity of interfacial ordering to amphiphilic molecules \cite{Bahr2006}. Toquer \textit{et al.} further demonstrated that molecular adsorption can directly control liquid crystal droplet shape, showing that interfacial composition influences both anchoring and overall droplet morphology \cite{Toquer2008}. More recently, Wei \textit{et al.} reported that molecular heterogeneity can drive reversible shape transformations in polymeric nematic liquid crystal droplets, emphasizing the coupling between interfacial organization and bulk liquid crystal elasticity \cite{Wei2019}. Together, these studies establish amphiphile adsorption as a powerful route for controlling liquid crystal droplet morphology through interfacial molecular organization.

The importance of interfacial effects is further supported by studies of liquid crystal interfacial tension. Previous work has shown that liquid crystal interfacial tension depends on temperature, orientational order, and interfacial molecular environment \cite{Wu2006,Harth2015}. In addition, microfluidic approaches have enabled quantitative measurements of interfacial tension and anisotropy in nematic systems \cite{Honaker2018,Honaker2021}. These studies suggest that small changes in interfacial composition can alter the balance between interfacial forces and liquid crystal elasticity, thereby modifying droplet reconfiguration pathways. At the same time, they indicate that interfacial tension alone may not fully capture liquid crystal interfacial behavior, since interfacial structure and orientational order are often strongly coupled.

Among thermotropic liquid crystals, cyanobiphenyls such as 4-octyl-4'-cyanobiphenyl (8CB) provide useful model systems for studying phase behavior, interfacial instabilities, and shape evolution. Earlier studies showed that 8CB droplets dispersed in aqueous surfactant environments can exhibit unusual breakup behavior and pronounced morphological changes near phase transitions \cite{Porter2012,Peddireddy2014Dissertation}. In the smectic regime, Peddireddy \textit{et al.} demonstrated the formation of myelin-like structures and extended smectic filaments at liquid crystal-aqueous interfaces \cite{Peddireddy2013Myelin,Peddireddy2014Dissertation}. These filamentous structures arise from the layered nature of the smectic phase and are relevant to both interfacial instabilities and applications such as optical waveguiding and lasing in smectic fibers \cite{Peddireddy2013OptExpress}. Related reports of fibrous self-assembly suggest that filament formation is a recurring outcome of coupling between mesophase order and interfacial organization \cite{Takenaka2022}.

A particularly important advance was reported by Peddireddy \textit{et al.}, who introduced a general strategy for self-shaping liquid crystal droplets by combining an amphiphile dissolved in the liquid crystal phase with a surfactant in the surrounding aqueous phase \cite{Peddireddy2021}. In that system, temperature ramps transformed initially spherical droplets into fibers that continuously changed in diameter. Their results showed that self-shaping behavior emerges from a balance between bulk liquid crystal elasticity and effective interfacial energy \cite{Peddireddy2021}. This framework provides a basis for understanding thermally driven droplet reconfiguration.

Despite these advances, most previous studies have focused on conventional ionic surfactants in the external aqueous phase. Comparatively little attention has been given to phospholipid-containing environments, even though lipids possess hydrocarbon chains that can introduce additional modes of interfacial organization relevant to biological interfaces \cite{Hamley2007,Buddingh2017}. This gap is significant because lipid molecular structure provides experimentally tunable parameters, including hydrocarbon-chain length, saturation, and conformational flexibility, that are known to influence self-assembly and shape transformations in other soft-matter systems \cite{Seifert1997,Sakashita2012,Ziherl2005}. However, it remains unclear whether these molecular characteristics can also regulate shape transformations in liquid crystal droplets.

In particular, hydrocarbon-chain saturation is expected to influence how lipids pack and organize at liquid crystal-aqueous interfaces. Saturated chains generally adopt more extended conformations and support denser molecular packing, whereas unsaturated chains possess cis-double-bond kinks that disrupt chain ordering. These differences can modify interfacial structure, anchoring conditions, and the transmission of elastic stresses between the interface and liquid crystal bulk. Consequently, comparing saturated and unsaturated phospholipids provides a route to investigate how molecular-scale chain conformation is translated into mesoscale droplet morphology. More broadly, such comparisons address a central question in responsive soft matter: how molecular organization at interfaces governs emergent shape transformations.

In this work, we investigate the temperature-dependent structural evolution of monoolein-doped 8CB droplets dispersed in aqueous phospholipid solutions. Monoolein acts as an internal amphiphile, while phospholipids in the aqueous phase populate the external interface. We specifically compare the saturated phospholipid 1,2-dilauroyl-\textit{sn}-glycero-3-phosphoglycerol (DLPG) and the unsaturated phospholipid 1,2-dioleoyl-\textit{sn}-glycero-3-phosphoglycerol (DOPG) to probe how hydrocarbon-chain saturation and conformational freedom influence thermally induced droplet reconfiguration. By varying both internal monoolein concentration and external phospholipid concentration, we examine how interfacial organization regulates droplet morphology during heating. Particular attention is given to the emergence of thin filaments in the smectic phase, the onset of more amorphous droplet deformation in the nematic phase, and transitions between smectic, nematic, and isotropic states.

Our results reveal a strong dependence of droplet morphodynamics on phospholipid chain architecture. DLPG-containing systems exhibit thin filamentation followed by an abrupt shape-change transition at the smectic-to-nematic phase transition, whereas DOPG-containing systems remain largely spherical and do not display shape-change behavior under the conditions examined. The contrast between these systems shows that shape change cannot be explained solely by interfacial tension and instead depends on how lipid molecular structure governs interfacial organization and its coupling to bulk liquid crystal elasticity. To our knowledge, this work provides the first observation of a discontinuous shape-change transition associated with a smectic-to-nematic phase transition. Through this approach, we identify how lipid-mediated interfacial organization, mesophase structure, and bulk liquid crystal elasticity together govern the thermal reconfiguration of liquid crystal droplets.

\section{Results and Discussion}

\subsection{Interfacial assembly and phase-dependent shape changes of 8CB/lipid droplets}

\begin{figure*}[t]
 \centering
    \centering
    \includegraphics[width=1\linewidth]{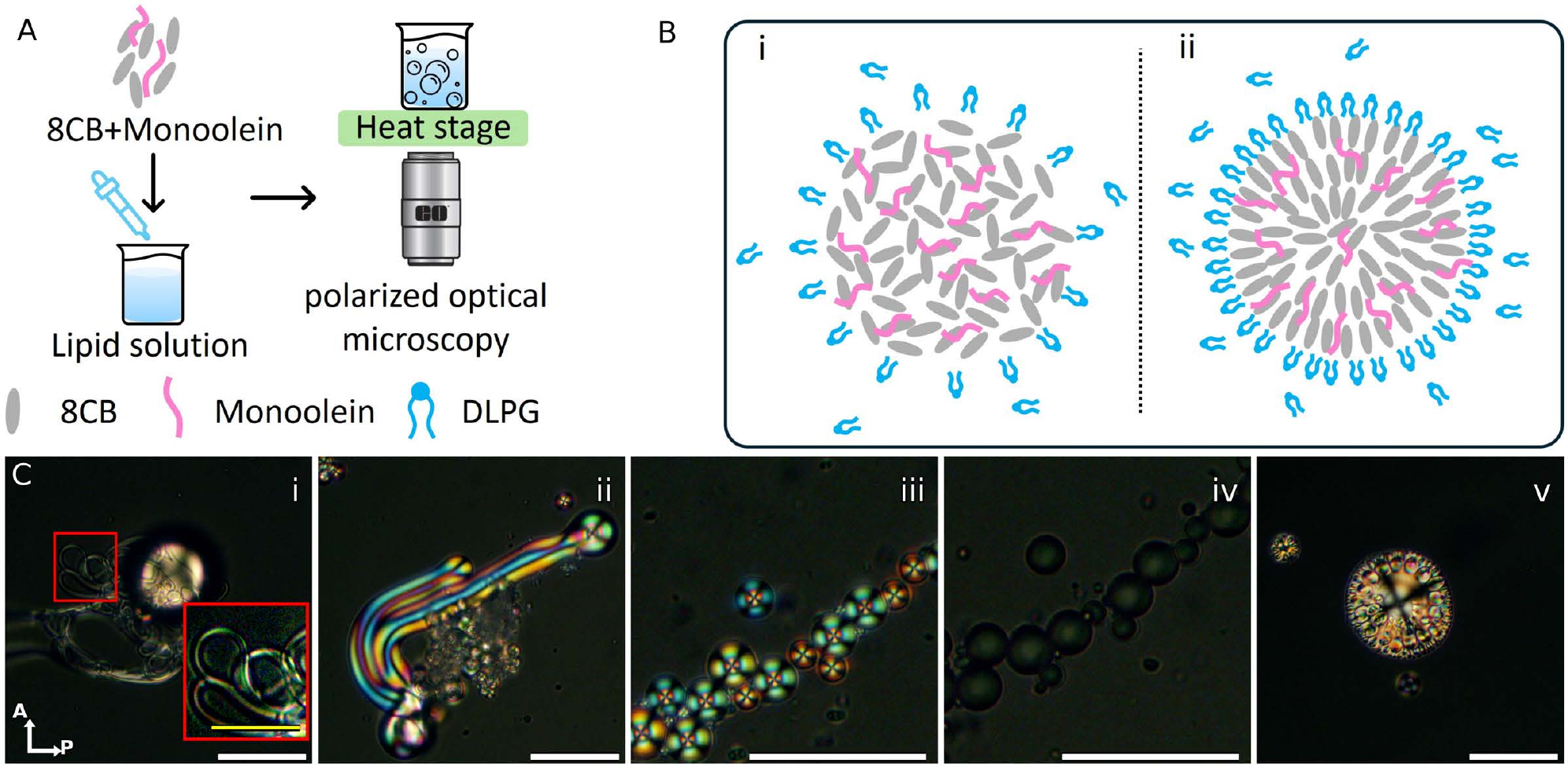}
    \caption{Experimental design, proposed interfacial organization, and representative structural states of monoolein-doped 8CB droplets in a DLPG-containing aqueous environment. (A) Schematic illustration of the experimental procedure. 8CB was doped with monoolein and dispersed in an aqueous lipid solution, and the droplets were observed during heating on a temperature-controlled stage by polarized optical microscopy. (B) Schematic illustration of the proposed interfacial organization of the droplets. (i) Amphiphilic molecules are proposed to be more loosely distributed at the liquid crystal-water interface. (ii) The interfacial amphiphiles are proposed to adopt a more ordered arrangement that promotes homeotropic anchoring. Gray denotes 8CB, magenta denotes monoolein, and blue denotes DLPG. (C) Representative polarized optical microscopy images showing the main structural states observed in this system: (i) thin filamentation, (ii) amorphous shape deformation, (iii) nematic state, (iv) isotropic state, and (v) smectic state. Scale bars: 100 μm (white). Scale bars: 50 μm (yellow).}
    \label{fig:setup}
\end{figure*}

To investigate how lipid-mediated interfacial organization influences liquid crystal droplet reconfiguration, 8CB was doped with monoolein (1-oleoyl-rac-glycerol) and dispersed in aqueous DLPG solutions. Following the strategy reported by Peddireddy \textit{et al.}, monoolein was introduced as an internal amphiphile to increase the coverage of surface-active molecules at the 8CB-water interface \cite{Peddireddy2021,Bahr2006}. The experimental design, proposed interfacial organization, and representative structural states are summarized in Fig.~1.

Fig.~1A illustrates the experimental workflow. Monoolein-doped 8CB droplets were dispersed in aqueous DLPG solutions and heated on a temperature-controlled microscope stage while being observed by polarized optical microscopy (POM, see Methods). This approach enabled simultaneous characterization of liquid crystal phase transitions, interfacial reorganization, and droplet morphology during thermal cycling \cite{LopezLeon2011,deGennesProst1995,Hamley2007}.

Fig.~1B shows the proposed interfacial organization. In one limiting case, amphiphilic molecules are distributed relatively loosely at the liquid crystal-water interface (Fig.~1B(i)). In the second, the amphiphiles adopt a more ordered arrangement that promotes homeotropic anchoring (Fig.~1B(ii)). The schematic is adapted from surfactant-mediated interfacial ordering models developed for liquid crystal droplets \cite{Peddireddy2021,Bahr2006}. Here, magenta denotes monoolein dissolved within the liquid crystal phase and blue denotes phospholipids in the surrounding aqueous phase.

The proposed interfacial structures provide a framework for interpreting the temperature-dependent droplet morphologies examined throughout this study. Because phospholipid packing depends strongly on hydrocarbon-chain architecture, differences in interfacial organization may determine whether thermally induced shape change occurs.

Representative structural states observed in the monoolein/DLPG system are shown in Fig.~1C. Thin birefringent filaments extending from the liquid crystal domain are shown in Fig.~1C(i), whereas the larger amorphous deformation state is shown in Fig.~1C(ii). Figures~1C(iii)-1C(v) show representative optical textures in undeformed droplets corresponding to the nematic, isotropic, and smectic phases, respectively. The nematic state exhibits birefringence characteristic of long-range orientational order (Fig.~1C(iii)), the isotropic state appears dark between crossed polarizers (Fig.~1C(iv)), and the smectic state displays a more complex birefringent texture (focal conic domains) consistent with layered order (Fig.~1C(v)) \cite{deGennesProst1995,Blanc2023,Neogi2026}. Interestingly, smectic droplets that exhibit focal conic domains are not found to form thin filaments. 

These structural states represent distinct manifestations of the coupling between mesophase structure and interfacial organization. Thin filaments are associated with the smectic state and resemble previously reported smectic myelin- and tube-like structures \cite{Peddireddy2012,Peddireddy2013Myelin,Peddireddy2014Dissertation}. By contrast, the larger deformed morphologies emerge after the smectic-to-nematic transition and are characterized by elongation, breakup, and non-spherical nematic domains rather than continued thin-filament growth. Distinguishing these two responses is essential because the results below show that thin smectic filamentation and thicker, more amorphous nematic deformation are separate phenomena associated with different states of molecular order.

\subsection{Shape changes from DLPG on 1 wt.-\% monoolein-doped 8CB droplets}

\begin{figure*}[t]
 \centering
    \centering
    \includegraphics[width=1\linewidth]{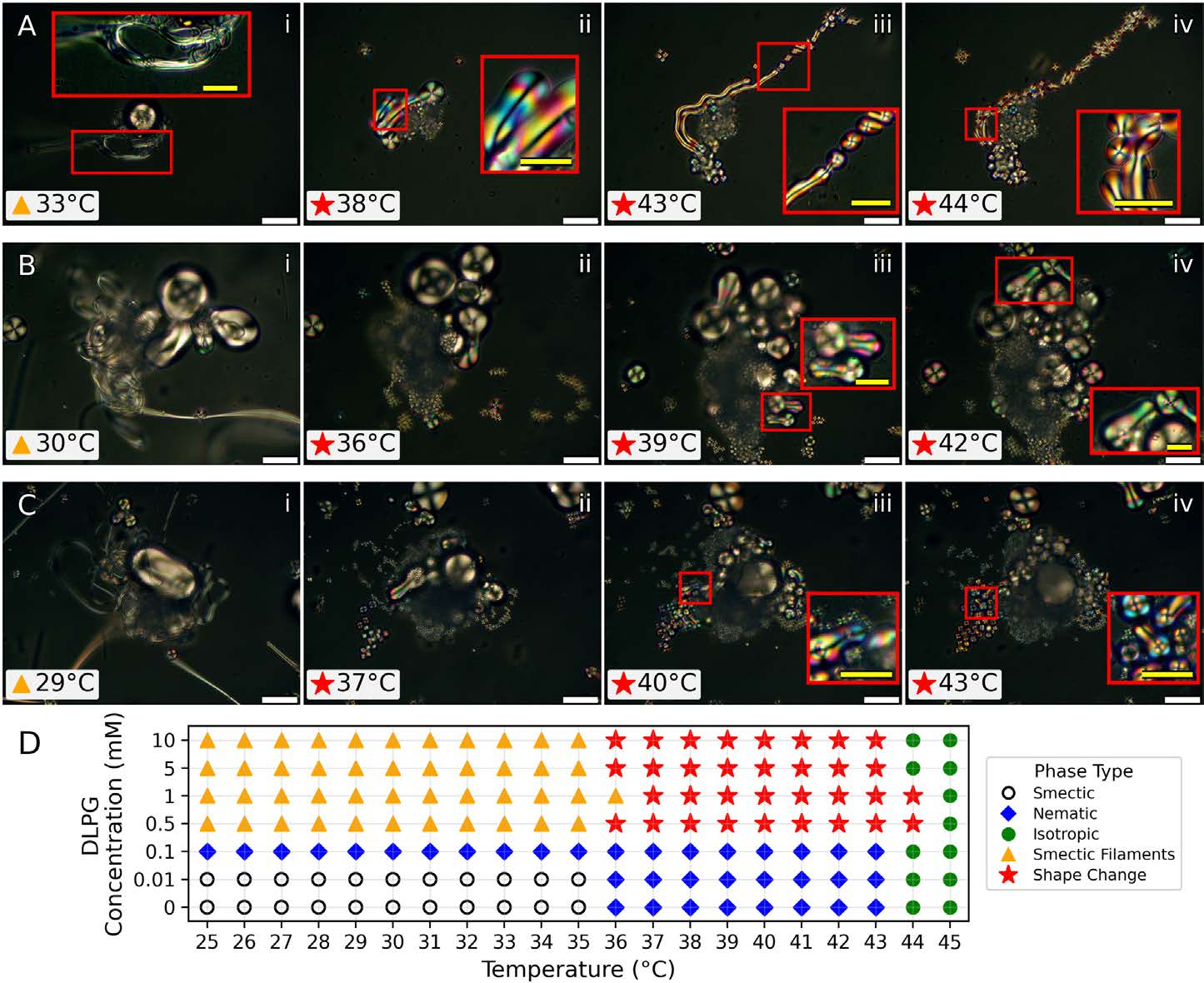}
    \caption{Temperature-dependent structural evolution of 1 wt.-\% monoolein-doped 8CB droplets in DLPG aqueous solutions. (A-C) Representative polarized optical microscopy images recorded during heating for droplets dispersed in aqueous solutions containing 1, 5, and 10 mM DLPG, respectively. The numbers in each panel indicate the temperature in $^\circ$C. The complete heating processes for 1 wt.-\% monoolein-doped 8CB droplets in 1, 5 and 10 mM DLPG solutions are shown in Movies S1, S2 and S3, respectively. The full video dataset (Movies S1-S9) is available online via the QR code in Fig.~S8 (DataverseNL, DOI: 10.34894/VXJ17C). The droplets underwent temperature-dependent transitions among smectic, nematic, and isotropic states, accompanied by filament formation and shape deformation. (D) Phase map summarizing the dominant structural states observed as a function of temperature and DLPG concentration. Scale bars: 100 μm (white). Scale bars: 50 μm (yellow).}
    \label{fig:dlpg1wt}
\end{figure*}

We first investigated the thermal reconfiguration of monoolein-doped 8CB droplets at a fixed monoolein concentration of 1 wt.-\% while varying the DLPG concentration from 0 to 10 mM. Representative results for 1, 5, and 10 mM DLPG are shown in Figs.~2A-2C, respectively, and the corresponding phase map is shown in Fig.~2D. Additional concentrations are provided in the Supporting Information (Figs.~S1 and S2). Samples were heated continuously from 25 to 45~$^\circ$C at 1~$^\circ$C min$^{-1}$ under polarized optical microscopy.

Across all DLPG concentrations, the droplets underwent the expected sequence of smectic, nematic, and isotropic phases upon heating. However, the accompanying morphological changes depended strongly on DLPG concentration (Fig.~2D). At low DLPG concentrations (0 and 0.01 mM), the droplets behaved similarly to bulk 8CB, remaining smectic at low temperature, becoming nematic near 36~$^\circ$C, and transitioning to the isotropic phase near 44~$^\circ$C (Figs.~S1 and S2). Increasing the DLPG concentration to 0.1 mM shifted the low-temperature behavior toward the nematic state, while 0.5 mM DLPG produced thin filaments between 25 and 35~$^\circ$C and a distinct shape-deformation state from 36~$^\circ$C onward.

These observations indicate that increasing DLPG concentration strengthens the coupling between interfacial organization and bulk liquid crystal ordering. The appearance of filamentation and shape change only above a threshold DLPG concentration suggests that droplet reconfiguration is not solely a consequence of intrinsic 8CB phase behavior. Instead, sufficient phospholipid accumulation and organization at the interface appear necessary before morphological instabilities develop.

In 1 mM DLPG (Fig.~2A), the droplets continuously ejected thin filaments throughout the smectic phase between 25 and 35~$^\circ$C. At 36~$^\circ$C, coincident with the smectic-to-nematic transition, the thin filaments rapidly retracted and the droplets entered a deformation regime characterized by elongation and breakup (Fig.~2A).
Upon reaching sufficiently large aspect ratios, the droplets fragmented into smaller daughter droplets that remained dynamically deformed throughout the nematic temperature range, similar to behavior reported previously \cite{Peddireddy2021}. The complete evolution is shown in Movie~S1.

The temporal sequence in Fig.~2A is particularly important. Thin filament growth occurs within the smectic regime, whereas larger droplet deformation appears only after the onset of the nematic phase. The abrupt retraction of thin filaments and the emergence of larger deformation at 36~$^\circ$C indicate a discontinuous shape-change transition coupled to the smectic-to-nematic phase transition.

At 5 mM DLPG (Fig.~2B), filament formation remained prominent throughout the smectic temperature range. As in the 1 mM system, filament retraction occurred near the smectic-to-nematic transition and was followed by droplet deformation. However, the resulting nematic morphologies were shorter and thicker than those observed at 1 mM DLPG (compare Figs.~2A(iii) and 2B(iii)). Some retracting filaments also fragmented into smaller droplets before fully reincorporating into the parent liquid crystal domain.

These observations show that increasing DLPG concentration alters the morphology of the shape-change state without eliminating the underlying sequence of smectic filamentation followed by nematic deformation. The reduced extent of elongation suggests that higher phospholipid concentrations modify how elastic stresses are redistributed during reconfiguration.

At 10 mM DLPG (Fig.~2C), the droplets displayed behavior qualitatively similar to that observed at 5 mM DLPG. Dense populations of thin filaments were present at low temperature (Fig.~2C(i)), followed by filament retraction near the smectic-to-nematic transition. Subsequent heating produced larger and more amorphous deformation events rather than highly elongated droplets (Figs.~2C(ii)-2C(iv)). The system remained in a dynamic non-spherical state throughout the nematic temperature range before reaching the isotropic phase, where the droplets became spherical again.

Fig.~2 establishes that DLPG concentration governs both the onset and pathway of thermally induced reconfiguration. At low concentrations, the droplets primarily follow the intrinsic phase behavior of 8CB. Above a threshold DLPG concentration, two distinct morphological regimes appear: a smectic regime characterized by thin filament growth and a nematic regime characterized by larger droplet deformation. The abrupt replacement of filamentation by deformation at the smectic-to-nematic transition provides the first indication that the shape-change transition is directly coupled to the underlying mesophase transition.

\subsection{Shape changes from DLPG on 2 wt.-\% monoolein-doped 8CB droplets}

\begin{figure*}[t]
 \centering
    \centering
    \includegraphics[width=1\linewidth]{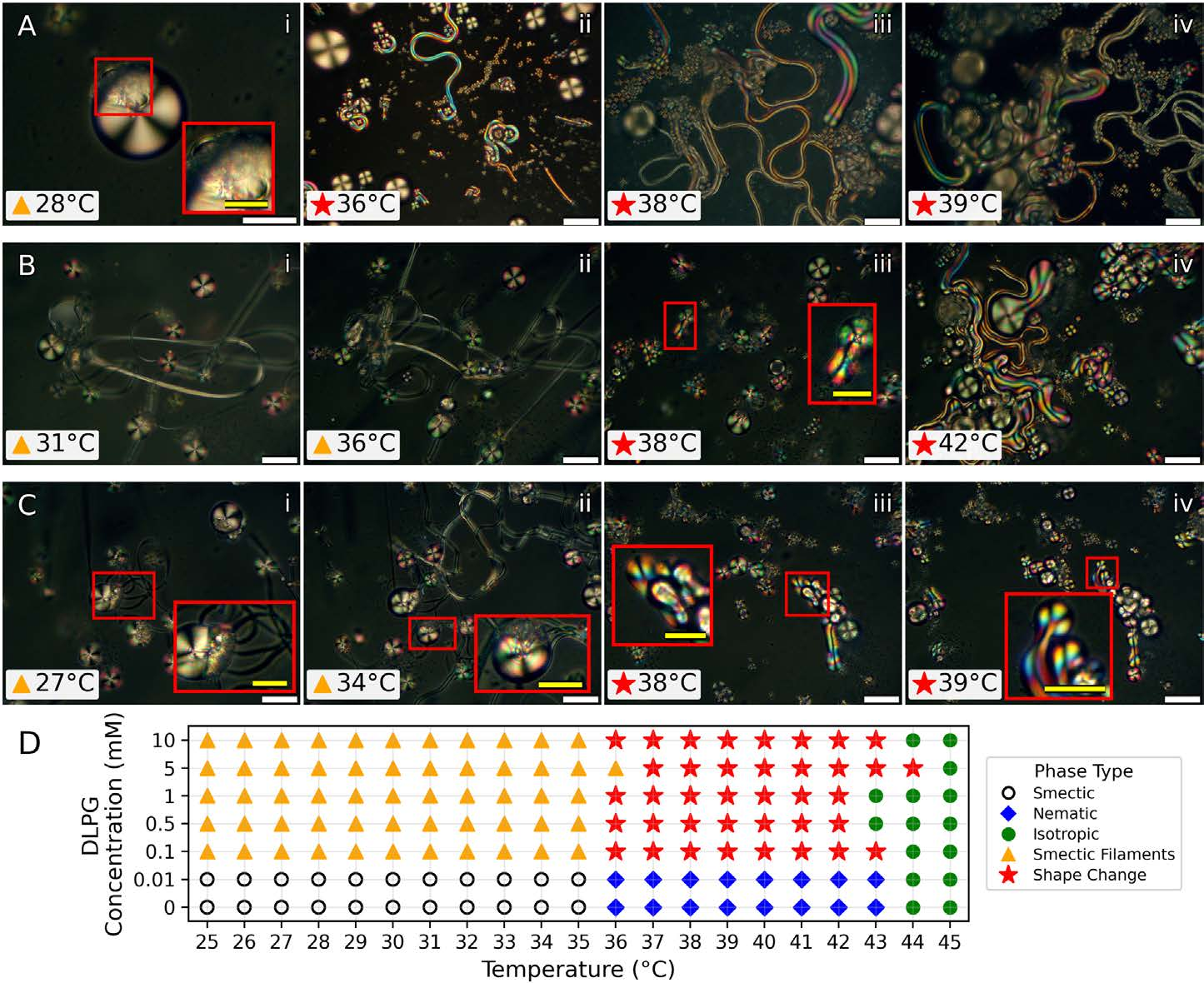}
    \caption{Temperature-dependent structural evolution of 2 wt.-\% monoolein-doped 8CB droplets in DLPG aqueous solutions. (A-C) Representative polarized optical microscopy images recorded during heating for droplets dispersed in aqueous solutions containing 1, 5, and 10 mM DLPG, respectively. The numbers in each panel indicate the temperature in $^\circ$C. The complete heating processes for 2 wt.-\% monoolein-doped 8CB droplets in 1, 5 and 10 mM DLPG solutions are shown in Movies S4, S5 and S6, respectively. The full video dataset (Movies S1-S9) is available online via the QR code in Fig.~S8 (DataverseNL, DOI: 10.34894/VXJ17C). The droplets underwent temperature-dependent transitions among smectic, nematic, and isotropic states, accompanied by filament formation and shape deformation. (D) Phase map summarizing the dominant structural states observed as a function of temperature and DLPG concentration. Scale bars: 100 μm (white). Scale bars: 50 μm (yellow).}
    \label{fig:dlpg2wt}
\end{figure*}

To further investigate the interplay between the internal amphiphile and the external lipid environment, the monoolein concentration was increased to 2 wt.-\%, while the DLPG concentration was varied from 0 to 10 mM. Representative micrographs for 1, 5, and 10 mM DLPG are shown in Figs.~3A-3C, respectively, and the corresponding phase map is shown in Fig.~3D. Additional data for 0, 0.01, 0.1, and 0.5 mM DLPG are provided in the Supporting Information (Figs.~S1 and S3). Relative to the 1 wt.-\% monoolein system (Fig.~2), increasing the internal monoolein concentration altered both smectic filamentation and subsequent nematic deformation.

At 1 mM DLPG (Fig.~3A), the droplets exhibited filament formation throughout the smectic temperature range. However, the filament density was lower than in the corresponding 1 wt.-\% monoolein system (compare Figs.~2 and 3, columns (i)). Upon heating through the smectic-to-nematic transition, the filaments retracted and the droplets entered a deformation regime characterized by elongated, thick fibers (Fig.~3A(ii)-3A(iv)). The deformed structures persisted throughout the nematic temperature range before becoming spherical in the isotropic state.

At 5 mM DLPG (Fig.~3B), both filament formation and subsequent deformation became more pronounced than in the 1 mM system. Smectic filaments were observed throughout the low-temperature regime and retracted near the smectic-to-nematic transition. After retraction, the droplets entered a deformation regime dominated by short, thick anisotropic structures that were less elongated than those observed at 1 mM DLPG (compare Figs.~3A(iii) and 3B(iii)).

At 10 mM DLPG (Fig.~3C), filament formation remained evident throughout the smectic phase. Upon heating into the nematic state, the droplets again underwent deformation, but the resulting structures remained relatively compact and showed limited elongation. Instead of producing highly anisotropic domains, the droplets predominantly formed shorter, thicker morphologies (Fig.~3C(iii)-3C(iv)).

The concentration dependence summarized in Fig.~3D demonstrates that both internal monoolein concentration and external DLPG concentration influence the reconfiguration pathway. Increasing monoolein modifies the balance between smectic filamentation and nematic deformation, while increasing DLPG changes the morphology of the deformed nematic state. These results further support the view that thermally induced droplet morphodynamics arise from coupling between interfacial organization and bulk liquid crystal elasticity, with internal and external amphiphiles jointly regulating that coupling.

\subsection{Shape changes from DLPG on 5 wt.-\% monoolein-doped 8CB droplets}

\begin{figure*}[t]
 \centering
    \centering
    \includegraphics[width=1\linewidth]{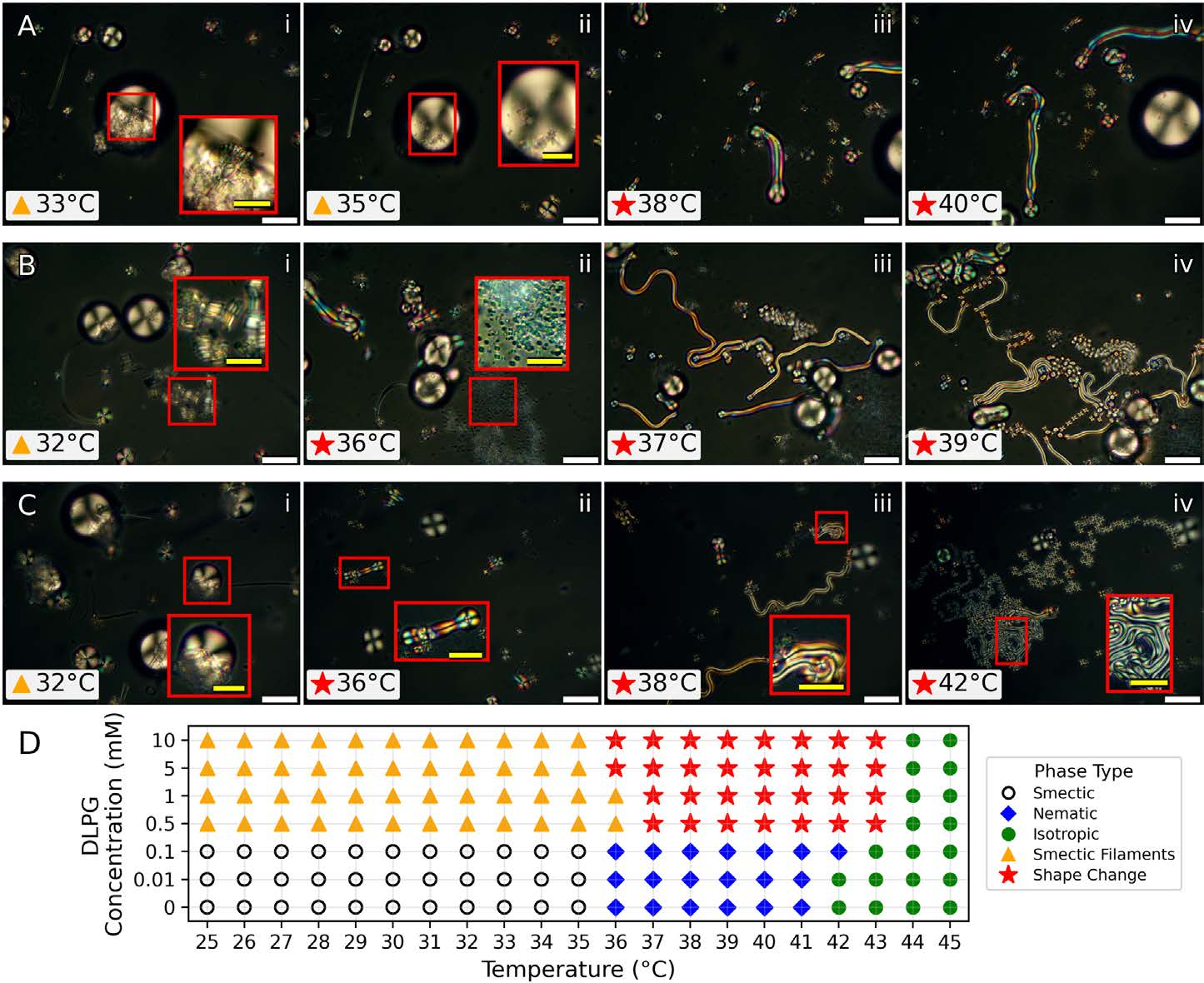}
    \caption{Temperature-dependent structural evolution of 5 wt.-\% monoolein-doped 8CB droplets in DLPG aqueous solutions. (A-C) Representative polarized optical microscopy images recorded during heating for droplets dispersed in aqueous solutions containing 1, 5, and 10 mM DLPG, respectively. The numbers in each panel indicate the temperature in $^\circ$C. The complete heating processes for 5 wt.-\% monoolein-doped 8CB droplets in 1, 5 and 10 mM DLPG solutions are shown in Movies S7, S8 and S9, respectively. The full video dataset (Movies S1-S9) is available online via the QR code in Fig.~S8 (DataverseNL, DOI: 10.34894/VXJ17C). The droplets underwent temperature-dependent transitions among smectic, nematic, and isotropic states, accompanied by filament formation and shape deformation. (D) Phase map summarizing the dominant structural states observed as a function of temperature and DLPG concentration. Scale bars: 100 μm (white). Scale bars: 50 μm (yellow).}
    \label{fig:dlpg5wt}
\end{figure*}

To investigate the effect of higher internal amphiphile loading, the monoolein concentration was increased to 5 wt.-\%. The same heating protocol and DLPG concentration range were used as in the previous experiments, with additional low-concentration data provided in the Supporting Information (Figs.~S1 and S4). Representative micrographs for 1, 5, and 10 mM DLPG are shown in Figs.~4A-4C, respectively, and the corresponding phase map is shown in Fig.~4D. Increasing the monoolein concentration to 5 wt.-\% produced a substantial shift in droplet morphodynamics relative to the 1 wt.-\% and 2 wt.-\% systems.

At 1 mM DLPG (Fig.~4A), the droplets exhibited smectic filament formation at low temperature followed by more amorphous deformation after the smectic-to-nematic transition. In contrast to the lower monoolein systems, however, the filamentous material tended to accumulate near the droplet surface rather than forming long, extended filaments (compare Figs.~2, 3, and 4, columns (i)). Upon further heating, the droplets adopted increasingly non-spherical morphologies before becoming isotropic.

At 5 mM DLPG (Fig.~4B), droplet restructuring became particularly pronounced. During the smectic phase, the droplets expelled large numbers of short, thin filaments (Fig.~4B(i)). These filaments folded, interacted with neighboring structures, and generated droplet motion (Movie~S8). Near the smectic-to-nematic transition, filament retraction occurred, but many filaments fragmented before being fully reabsorbed into the parent droplets. Above 36~$^\circ$C, droplet deformation became the dominant mode of structural evolution, producing persistent non-spherical nematic fibers (Fig.~4B).

At 10 mM DLPG (Fig.~4C), filament formation was again observed throughout the smectic regime (Fig.~4C(i)). Similar to the 1 mM condition, much of the filamentous material remained localized near the droplet surface. Following the smectic-to-nematic transition, the droplets entered a deformation regime, although the resulting structures remained relatively short and showed thinner nematic fibers before the isotropic transition (Fig.~4C(iii)-4C(iv)).

These observations reveal a systematic evolution of droplet behavior with increasing monoolein concentration. Compared with the 1 wt.-\% and 2 wt.-\% systems, the 5 wt.-\% droplets show a reduced tendency to generate long, straight smectic filaments and a greater tendency to retain filamentous material near the droplet interface. The expelled smectic filaments also become increasingly folded and curved, suggesting that monoolein modifies the elastic properties of the smectic phase and lowers the energetic penalty for filament bending. However, the effect of monoolein is not limited to the smectic regime. After the droplets enter the nematic phase, higher monoolein concentrations promote more persistent elongated deformations that retain a filament-like character over a broader temperature and DLPG concentration range. Thus, increasing monoolein concentration does not eliminate the distinction between smectic filamentation and nematic deformation, but changes how each response is expressed in its respective mesophase.

Taken together, Figs.~2-4 show that internal and external amphiphiles regulate different aspects of droplet morphodynamics. The internal monoolein concentration affects both the morphology of the smectic filaments and the persistence of elongated structures in the nematic deformation regime. In the smectic phase, increasing monoolein suppresses long, straight filaments while promoting filament bending, folding, and retention of filamentous material near the droplet interface. In the nematic phase, monoolein appears to stabilize elongated shape-change structures, possibly by modifying the elastic properties. By contrast, the external DLPG concentration primarily modulates the onset and character of thermally induced restructuring, with higher DLPG concentrations tending to favor thicker, shorter, and more fragmented nematic shape-change structures. Across all conditions examined, thin filament formation remains associated with the smectic phase, whereas thicker and more amorphous deformations remain associated with the nematic phase, with a discontinuous shape-change transition at the smectic-to-nematic boundary.

Comparison experiments performed in DOPG and mixed DLPG solutions provide additional insight into the role of lipid molecular structure (Figs.~S5 and S6). In DOPG-only solutions, the droplets largely retained spherical morphologies and displayed conventional nematic textures, with little evidence of either smectic filamentation or nematic droplet deformation. Mixed DLPG systems exhibited intermediate behavior, with some shape-change structures present but considerably weaker than in DLPG-only solutions.

These observations establish that filamentation and shape change depend strongly on phospholipid identity. Because DLPG and DOPG possess identical headgroups but differ substantially in hydrocarbon-chain structure, the results point to chain saturation and conformational freedom as critical factors in liquid crystal droplet morphodynamics. Saturated DLPG promotes the interfacial organization required for filament formation and discontinuous shape change, whereas conformationally less-flexible unsaturated DOPG does not produce comparable behavior under otherwise similar conditions.

\subsection{Interfacial tension of 5CB in DLPG and DOPG aqueous solutions}

\begin{figure}[t]
 \centering
    \includegraphics[width=1\linewidth]{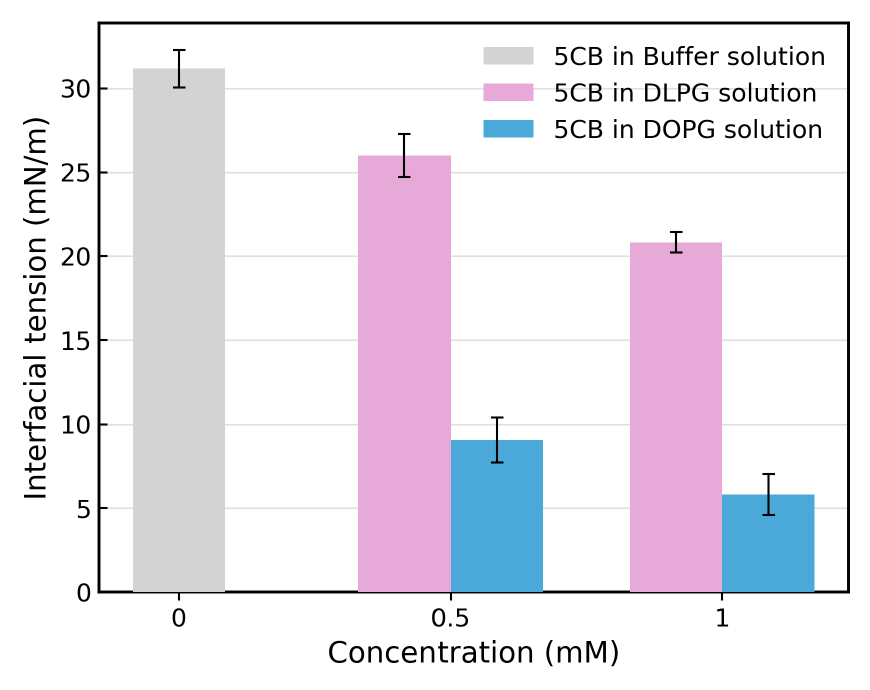}
   \caption{Interfacial tension of 5CB in buffer, DLPG, and DOPG aqueous solutions at different lipid concentrations. The interfacial tension was measured for 5CB droplets in buffer solution and in aqueous DLPG and DOPG solutions containing 0.5 and 1 mM lipid. The gray bar represents 5CB in buffer solution, the pink bars represent 5CB in DLPG solution, and the blue bars represent 5CB in DOPG solution. Error bars represent standard deviations.}
    \label{fig:ift}
\end{figure}

To determine whether the distinct morphologies observed in DLPG- and DOPG-containing systems could be explained by differences in interfacial activity, we measured the interfacial tension of 4-cyano-4'-pentylbiphenyl (5CB), used here as a proxy for 8CB (which is extremely viscous at room temperature), in buffer and in aqueous DLPG and DOPG solutions containing 0.5 and 1 mM lipid. The results are summarized in Fig.~5.

In the absence of added lipid (10 mM Tris buffer with 0.1 M NaCl at pH 9.68, see Methods), the interfacial tension of the 5CB-water interface was approximately 31 mN/m, consistent with previous measurements of 5CB-water interfaces \cite{Honaker2021}. Addition of either DLPG or DOPG reduced the interfacial tension in a concentration-dependent manner, indicating adsorption of phospholipids at the liquid crystal interface. However, the magnitude of the reduction differed substantially between the two lipids. In DLPG solutions, the interfacial tension decreased to approximately 26 mN/m at 0.5 mM and 21 mN/m at 1 mM. By contrast, DOPG produced a much larger reduction, decreasing the interfacial tension to approximately 9 mN/m at 0.5 mM and 6 mN/m at 1 mM.

These measurements show that DOPG is considerably more effective than DLPG at lowering the liquid crystal-aqueous interfacial tension. Because DLPG and DOPG possess the same phosphoglycerol headgroup but differ in hydrocarbon-chain structure, this difference is likely related to their distinct packing behavior. DOPG contains two cis-unsaturated oleoyl chains that are kinked at the unsaturated bond and less conformationally flexible, whereas DLPG contains two saturated lauroyl chains capable of adopting more extended conformations and efficient packing (Fig.~S7). Although the present measurements do not directly resolve molecular organization at the interface, they demonstrate that hydrocarbon-chain architecture strongly influences liquid crystal-aqueous interfacial properties.

Importantly, the interfacial tension trends do not mirror the droplet morphologies. Despite producing the largest reduction in interfacial tension, DOPG does not induce the extensive smectic filamentation or discontinuous shape-change transition observed in DLPG-containing systems. Instead, DOPG-containing droplets remain largely spherical and display comparatively conventional phase behavior (Fig.~S5). Mixed DLPG:DOPG systems exhibit intermediate responses (Fig.~S6), consistent with partial suppression of the DLPG-induced restructuring pathway.

This contrast shows that interfacial tension alone does not govern droplet morphodynamics. If interfacial tension were the dominant control parameter, DOPG would be expected to produce the strongest shape transformations. The experiments show the opposite trend. Therefore, the emergence of smectic filamentation and discontinuous shape change must depend on additional factors, including the molecular organization of adsorbed lipids, the anchoring conditions they generate, and their coupling to bulk liquid crystal elasticity and mesophase structure.

Overall, the interfacial-tension measurements and droplet observations identify hydrocarbon-chain architecture as a key control parameter for thermally induced liquid crystal droplet reconfiguration. Saturated DLPG supports interfacial conditions that enable filament formation and shape change, whereas kinked, unsaturated DOPG strongly lowers interfacial tension but does not produce comparable morphologies. These findings support the central hypothesis of this work: molecular-scale differences in lipid-chain saturation and conformational freedom regulate how interfacial organization is translated into mesoscale shape transformations.

\subsection{Phenomenological description of nematic self-shaping}

Above the smectic-nematic transition temperature $T_{SN}$, the layered structure disappears and the dominant morphology becomes droplet deformation, elongation, and fragmentation. We interpret this regime using the phenomenological equilibrium model proposed by Peddireddy \textit{et al.} \cite{Peddireddy2021}, which builds on the escaped-radial director configuration analyzed by Cladis and Kleman \cite{Cladis1972} and independently by Meyer \cite{Meyer1973}.

For a cylindrical nematic fiber with homeotropic surface anchoring with radius $r_f$, a singular radial director field is avoided by escape of the director along the fiber axis. The minimum-energy director configuration is
\begin{equation}
\mathbf{n}(r)
=
\cos\varphi(r),\hat{\mathbf r}
+
\sin\varphi(r),\hat{\mathbf z},
\end{equation}
where $\varphi(r_f)=0$ at the surface and $\varphi(0)=\pi/2$ at the fiber center. Following Cladis and Kleman, the elastic energy per unit length of this escaped-radial configuration is
\begin{equation}
\frac{F}{L}
=
\Omega\pi K_{11},
\end{equation}
with
\begin{equation}
\Omega
=
2+
\beta
\frac{\arctan\sqrt{\beta-1}}
{\sqrt{\beta-1}},
\qquad
\beta
=
\frac{K_{33}}{K_{11}}.
\end{equation}
The equilibrium morphology results from competition between elastic distortions and interfacial free energy,
\begin{equation}
F
=
F_{\rm N}
+
\gamma_{\rm eff}A,
\end{equation}
where
\begin{equation}
\begin{aligned}
F_{\rm N}
=
\frac{1}{2}\int
\Big[
&K_{11}(\nabla\cdot\mathbf n)^2 \\
&+K_{22}(\mathbf n\cdot\nabla\times\mathbf n)^2 \\
&+K_{33}(\mathbf n\times\nabla\times\mathbf n)^2
\Big]
\,dV.
\end{aligned}
\end{equation}
and $A$ is the total liquid crystal-aqueous interfacial area. Minimization of the Gibbs free energy under volume conservation yields an equilibrium fiber radius \cite{Peddireddy2021}
\begin{equation}
r_f
=
\frac{\Omega K_{11}}
{\gamma_{\rm eff}}.
\end{equation}
Thus, the characteristic dimensions of the shape-change structures are controlled by two quantities: the elastic anisotropy encoded in $\Omega(\beta)$ and the effective interfacial free energy $\gamma_{\rm eff}$.

The amorphous and fragmented structures observed in DLPG solutions are broadly consistent with this framework. Unlike the long escaped-radial fibers reported by Peddireddy \textit{et al.} for cetyltrimethylammonium bromide (CTAB)/monoolein systems, increasing DLPG concentration produces progressively shorter and thicker structures. Pendant-drop measurements show that DLPG lowers the liquid crystal-aqueous interfacial tension, thereby reducing the interfacial penalty associated with deformation. As the magnitude of $\gamma_{\rm eff}$ decreases, the energetic cost of generating additional interface becomes smaller, favoring breakup into multiple shorter and more compact domains rather than a single elongated fiber.

The experiments further indicate that monoolein affects the elastic contribution to the free energy. Although increasing monoolein suppresses the formation of long smectic filaments below $T_{SN}$, the resulting nematic structures remain elongated over a broader temperature and DLPG concentration range and retain a more fiber-like character. Because monoolein is dissolved directly within the liquid-crystal phase, it can modify local orientational correlations and the elastic anisotropy parameter $\beta=K_{33}/K_{11}$. One possibility is that the oleyl chain of monoolein perturbs pretransitional smectic fluctuations and suppresses the anomalous increase in $K_{33}$ that normally occurs near $T_{SN}$. A reduction in $K_{33}$ relative to $K_{11}$ decreases $\beta$ and stabilizes the escaped-radial configuration, thereby promoting the persistence of elongated nematic structures.

Within this picture, DLPG and monoolein influence different terms in the free energy. DLPG primarily modifies $\gamma_{\rm eff}$ through lipid adsorption and interfacial organization, whereas monoolein affects the elastic response encoded in $\Omega(\beta)$. Increasing DLPG concentration favors shorter, thicker, and more fragmented structures, while increasing monoolein promotes the persistence of elongated nematic deformations. The observed morphologies therefore arise from the competition between interfacial effects governed by the external lipid environment and elastic effects controlled by the internal amphiphile.

Comparison with DOPG further demonstrates that interfacial tension alone cannot account for the observed behavior. Although DOPG lowers the liquid crystal-aqueous interfacial tension substantially more than DLPG, it does not produce either smectic filamentation or the nematic shape change. We therefore propose that the saturated lauroyl chains of DLPG form a more ordered interfacial layer that more effectively transmits anchoring constraints and elastic stresses into the bulk liquid crystal, whereas the conformational kinks associated with the cis-unsaturated oleoyl chains of DOPG generates a more fluid interfacial layer that lowers interfacial tension efficiently but couples less strongly to bulk orientational order.

\subsection{Phenomenological description of smectic filament formation}

Below the smectic-nematic transition temperature $T_{SN}$, the dominant morphological response is the formation of thin, birefringent filamentous structures that emerge from the droplet surface and persist throughout the smectic phase.

The smectic filaments observed here resemble the myelin-like structures reported previously in smectic liquid-crystal systems and at liquid crystal-aqueous interfaces, where the liquid crystal was not doped with monoolein \cite{Peddireddy2013Myelin,Peddireddy2014Dissertation}. The smectic-A free energy may be written in terms of the layer displacement field $u(\mathbf r)$ as
\begin{equation}
F_{\mathrm{SmA}}
=
\int
\left[
\frac{B}{2}
\left(
\frac{\partial u}{\partial z}
\right)^2
+
\frac{K}{2}
\left(
\nabla_{\perp}^{2}u
\right)^2
\right]
dV
+
\gamma_{\mathrm{eff}}A,
\end{equation}
where $B$ and $K$ are the layer-compression and layer-bending moduli, respectively, $A$ is the liquid crystal-aqueous interfacial area, and $\gamma_{\mathrm{eff}}$ is the effective interfacial free-energy density.

For a cylindrical stack of curved smectic layers, the dominant elastic contribution arises from layer bending. Following the treatment of curved smectic layers developed by Williams and Kleman \cite{Kleman1975,Oswald2005} (and assuming minimal layer compression), the layer-curvature energy of a filament of length $L$ and radius $r_\text{sm}$ scales as
\begin{equation}
F_{\mathrm{layer}}
\approx
\pi K L
\ln\left(\frac{r_\text{sm}}{a}\right),
\end{equation}
where $a$ is a molecular-scale cutoff associated with the filament core. The total filament free energy is then
\begin{equation}
F_\text{filament}
\approx
\pi K L
\ln\left(\frac{r_\text{sm}}{a}\right)
+
2\pi r_\text{sm} L \gamma_{\mathrm{eff}},
\end{equation}
which balances the elastic cost of maintaining curved layers against the interfacial free-energy gain associated with amphiphile adsorption. Minimization with respect to $r_\text{sm}$ gives
\begin{equation}
r_\text{sm}
\approx
-\frac{K}{2\gamma_{\mathrm{eff}}},
\end{equation}
analogous to the radius selection obtained by Peddireddy \textit{et al.} for escaped-radial nematic fibers.

Using the characteristic diameter of the nematic structures ($\sim50~$μm) together with representative Frank elastic constants for cyanobiphenyl liquid crystals ($K_{11}\approx7$ pN) \cite{Madhusudana1982,deGennesProst1995} gives an effective interfacial free energy of order $|\gamma_{\mathrm{eff}}|\sim10^{-6}$ N,m$^{-1}$. Combining this estimate with the observed smectic filament diameter of approximately $1~$μm yields an effective layer-bending modulus of order $K\sim10^{-12}$ N, roughly one order of magnitude smaller than reported values for bulk smectic 8CB. This reduction suggests that monoolein substantially softens the smectic phase.

Experimentally, increasing monoolein concentration primarily changes the morphology rather than the existence of the smectic filaments. At low monoolein content, relatively long and straight filaments are produced, whereas higher monoolein concentrations lead to more folded and highly bent structures that remain close to the droplet surface. We attribute this behavior to the disruption of smectic order by monoolein, which reduces the effective layer-bending modulus and thereby lowers the energetic penalty for filament curvature.

The comparison between DLPG and DOPG further demonstrates that bulk smectic order alone is insufficient to generate filamentation. Although DOPG-containing droplets still undergo the smectic phase transition, they do not form filaments. Thus, filamentation depends not only on smectic ordering but also on the interfacial free-energy balance represented by $\gamma_{\mathrm{eff}}$.

A possible origin of this difference lies in the hydrocarbon-chain architecture of the phospholipids. DLPG contains saturated lauroyl chains that can adopt relatively extended conformations and pack efficiently at the interface, promoting stronger coupling between interfacial organization and the adjacent smectic layers. By contrast, the cis-unsaturated oleoyl chains of DOPG introduce conformational disorder and may form a more fluid interfacial layer that lowers interfacial tension efficiently but couples less strongly to smectic elasticity. Consequently, smectic order exists in both systems, but only DLPG stabilizes filament formation over an experimentally accessible range of temperatures and compositions.

Interestingly, the absence of filament formation in smectic droplets containing focal conic domains further suggests that filamentation and focal conic domain formation are competing modes of elastic relaxation. In the filament-forming pathway, amphiphile-mediated lowering of the effective interfacial free energy allows the droplet to accommodate smectic elasticity by deforming the interface and generating thin cylindrical extensions. In the non-filamenting pathway, interfacial deformation is not sufficiently favorable, and the smectic phase instead relaxes through internal defect structures such as focal conic domains. Thus, the absence of filaments in droplets containing focal conic domains further indicates that filamentation is not simply a consequence of smectic order, but requires an interfacial pathway that makes shape deformation energetically favorable \cite{Cladis1972}.

\subsection{Discontinuous shape-change transition and the role of lipid architecture}

The abrupt onset of droplet deformation at $T_{SN}$ can be understood by considering the two phenomenological descriptions together. Below $T_{SN}$, the droplet morphology is governed by smectic layer elasticity and interfacial organization; above $T_{SN}$, the layered structure is lost and the relevant morphology is governed by nematic Frank elasticity and the escaped-radial self-shaping mechanism described by Peddireddy \textit{et al.} \cite{Peddireddy2021}. The observed transition therefore reflects a switch between two distinct shape-selection mechanisms rather than a continuous evolution of a single filamentous structure.

In the smectic phase, a cylindrical filament is expected to contain curved or concentric smectic layers. Because the smectic layer normal is coupled to the director, this geometry imposes a radial director field, $\mathbf n=\hat{\mathbf r}$, together with a singular core or defect-like region along the filament axis. This structure is stabilized only when smectic layering is present and when the interfacial free-energy gain associated with DLPG-mediated organization compensates for the elastic cost of curved layers. Upon heating through $T_{SN}$, long-range smectic order is lost and the layer-compression constraint that maintains this radial smectic geometry disappears. The thin smectic filaments are therefore no longer supported by the smectic elasticity and rapidly retract into the parent droplets.

Above $T_{SN}$, the same homeotropic boundary condition no longer requires purely radial director. Instead, the nematic can lower its elastic energy through the escaped-radial configuration of Cladis and Kleman, in which the director develops an axial component and avoids a singular line defect \cite{Cladis1972}. The nematic deformation regime is therefore governed by a different free-energy balance, involving Frank elasticity and the effective interfacial free energy. The discontinuous shape change observed at $T_{SN}$ is thus interpreted as an exchange of stability between a smectic filament state and a nematic deformation state. This interpretation is consistent with the large difference in characteristic length scales observed experimentally. Thin smectic filaments are replaced by much thicker and more amorphous nematic deformations.

This pathway differs from that reported by Peddireddy \textit{et al.} for conventional CTAB-monoolein systems \cite{Peddireddy2021}. In those systems, the dominant self-shaping process occurs in the nematic phase, where escaped-radial fibers grow and evolve during cooling. At the nematic-to-smectic transition, the escaped-radial configuration becomes incompatible with smectic layering and the fibers either break into droplets or, under conditions that suppress breakup, transform into smectic fibers with continuous changes in fiber diameter across the N-SmA phase transition. In contrast, the DLPG system studied here exhibits thin smectic filaments before the transition, followed by abrupt filament retraction and the onset of a thicker nematic deformation regime upon heating through $T_{SN}$. The lipid system therefore follows a distinct morphological pathway in which thin smectic filamentation precedes nematic self-shaping rather than emerging from it.

The difference between these pathways highlights the role of interfacial amphiphile architecture. DLPG is a double-chain, saturated phospholipid, whereas CTAB is a single-chain surfactant. The saturated double-chain structure of DLPG is expected to support a more ordered and densely packed interfacial layer, which may couple more strongly to smectic layering at the liquid crystal-aqueous interface. This coupling can help stabilize thin smectic filaments below $T_{SN}$ by allowing interfacial organization to interact with the bulk smectic. By contrast, single-chain surfactants such as CTAB can support homeotropic anchoring sufficient for nematic escaped-radial fiber formation, but appear less effective at stabilizing the smectic filament state observed here.

The comparison with DOPG further supports this interpretation. Although DOPG is also a double-chain phospholipid, its cis-unsaturated oleoyl chains introduce conformational disorder and likely form a more fluid interfacial layer. Thus, DOPG lowers the measured interfacial tension efficiently but does not stabilize smectic filaments or produce the discontinuous shape-change transition observed with DLPG. Together, these comparisons indicate that the relevant control parameter is not simply amphiphile adsorption or surface activity, but the way hydrocarbon-chain architecture determines interfacial order and its coupling to bulk liquid crystal phase behavior. More broadly, the results suggest that single-chain surfactants, saturated phospholipids, and unsaturated phospholipids can select distinct morphological pathways during thermally induced liquid crystal reconfiguration.

\section{Conclusion}

In this work, we investigated the thermally induced structural evolution of monoolein-doped 8CB droplets dispersed in aqueous phospholipid solutions. By combining polarized optical microscopy with interfacial-tension measurements, we examined how internal amphiphiles and external lipid environments regulate the coupling between interfacial organization, liquid crystal phase behavior, and droplet morphology.

The droplets exhibited temperature-dependent structural evolution during heating, including smectic, nematic, and isotropic phase transitions accompanied by filament formation and droplet deformation. Across DLPG-containing systems, two distinct morphological regimes were identified. In the smectic phase, the droplets generated thin, birefringent filamentous structures. Upon heating through the smectic-to-nematic transition, the filaments retracted and were replaced by a deformation regime characterized by elongation, breakup, and persistent non-spherical morphologies. The abrupt onset of deformation at this transition demonstrates that smectic filamentation and droplet shape change are distinct phenomena associated with different mesophase states. To our knowledge, this study provides the first observation of a discontinuous droplet shape-change transition coupled directly to a smectic-to-nematic phase transition.

Systematic variation of DLPG and monoolein concentrations showed that thermally induced droplet morphodynamics are jointly controlled by the external lipid environment and internal amphiphile content. Increasing DLPG concentration promoted the emergence of filamentation and nematic shape-change behavior, while increasing monoolein concentration altered both the smectic and nematic responses. In the smectic phase, higher monoolein concentrations produced shorter, more folded, and more highly curved filaments, suggesting a reduction in the effective elastic penalty for smectic-layer bending. In the nematic phase, higher monoolein concentrations promoted more persistent elongated deformations that retained a filament-like character over a broader temperature and DLPG concentration range. These observations suggest that monoolein modifies the elastic properties of both mesophase regimes, whereas DLPG primarily regulates the interfacial conditions that enable thermally induced reconfiguration. Thus, droplet reconfiguration is governed not only by the intrinsic phase behavior of 8CB but also by the combined effects of internal amphiphile-mediated elasticity and external lipid-mediated interfacial organization.

The phenomenological framework developed here suggests that the observed morphologies arise from two distinct free-energy landscapes separated by the smectic-to-nematic phase transition. Below $T_{SN}$, filament formation is governed by competition between smectic elastic energies and interfacial free energy. In this regime, increasing monoolein concentration is proposed to disrupt smectic ordering and reduce the effective smectic elastic constants, lowering the energetic penalty for filament bending and promoting folded morphologies. Above $T_{SN}$, where layered order is lost, droplet morphology is instead governed by the balance between Frank elasticity and interfacial free energy, consistent with the self-shaping model of Peddireddy \textit{et al.}. In this regime, increasing DLPG concentration modifies the effective interfacial conditions that control droplet deformation, fragmentation, and the characteristic dimensions of the shape-change structures.

Comparison of DLPG and DOPG further revealed a pronounced dependence on phospholipid hydrocarbon-chain architecture. DLPG-containing systems exhibited extensive smectic filamentation that transitioned discontinuously (with filament retraction) at the liquid crystal phase-transition temperature to a more amorphous shape-change behavior, whereas DOPG-containing systems remained largely spherical and showed little comparable restructuring. Importantly, droplets in DOPG solutions still exhibited smectic ordering but did not generate extended filaments, demonstrating that smectic order alone is not sufficient for filamentation. Interfacial-tension measurements showed that DOPG reduced the liquid crystal-aqueous interfacial tension substantially more than DLPG, yet did not induce filamentation or shape change. This contrast shows that interfacial tension alone cannot explain the observed morphologies. Rather, hydrocarbon-chain number, saturation, length, and conformational freedom influence how phospholipids organize at liquid crystal interfaces and how that organization couples to anchoring, elastic stresses, and phase transitions within the liquid crystal.

Additional insight emerges from comparison with self-shaping liquid crystal systems based on conventional single-chain amphiphiles such as CTAB \cite{Peddireddy2021}. Although CTAB-containing systems exhibit temperature-dependent droplet reconfiguration, they do not exhibit the abrupt shape-change transition observed here at the smectic-to-nematic phase boundary. The appearance of a discontinuous shape-change transition in DLPG-containing systems suggests that phospholipid molecular architecture introduces an additional coupling between interfacial organization and bulk liquid crystal ordering. A possible explanation is that the saturated double-chain structure of DLPG promotes a more ordered interfacial layer that couples strongly to smectic order at the liquid crystal interface. Such coupling may stabilize smectic filaments below $T_{SN}$ and create the conditions for abrupt morphological reconfiguration when layered order disappears. By contrast, single-chain surfactants and unsaturated phospholipids may form more fluid interfacial layers that couple less strongly to smectic ordering. Future studies that systematically vary saturated hydrocarbon-chain length could further clarify how molecular packing at interfaces controls the emergence of mesoscale liquid crystal morphologies.

Overall, these findings establish lipid molecular structure as an important control parameter for liquid crystal droplet morphodynamics. More broadly, they demonstrate how molecular-scale differences in hydrocarbon-chain conformation and packing can be translated into mesoscale shape transformations through coupling between interfacial assembly and liquid crystal ordering. The results further suggest that lipid-dependent interfacial organization governs not only the magnitude of interfacial forces but also how those forces are transmitted to the bulk liquid crystal. This framework provides new insight into responsive liquid crystal systems and offers a basis for designing reconfigurable soft materials whose morphology is controlled through molecular organization at interfaces.

\section{Materials and Methods}
\subsection{Materials}

The lipids 1,2-dilauroyl-\textit{sn}-glycero-3-phosphoglycerol (DLPG) and 1,2-dioleoyl-\textit{sn}-glycero-3-phosphoglycerol (DOPG) were obtained from Avanti Polar Lipids. Monoolein (1-oleoyl-\textit{rac}-glycerol), Tris buffer powder, and chloroform were purchased from Sigma-Aldrich. Chloroform (C2432, $\geq$99.5\%, stabilized with 100-200 ppm amylenes) was used as the organic solvent for preparing lipid films and monoolein stock solutions. The aqueous phase consisted of 10 mM Tris and 0.1 M NaCl at pH 9.68. The liquid crystals 4-cyano-4'-pentylbiphenyl (5CB) and 4-octyl-4'-cyanobiphenyl (8CB) were purchased from TCI Chemicals. 

\subsection{Liposome Preparation}
Lipid vesicles were prepared using the standard thin-film hydration method. DLPG or DOPG was first dissolved in chloroform in a glass vial. The organic solvent was then evaporated under a gentle stream of nitrogen gas to form a uniform thin lipid film on the vial wall. To ensure complete removal of residual solvent, the lipid film was kept under vacuum overnight. The dried film was subsequently hydrated with 10 mM Tris buffer containing 0.1 M NaCl (pH 9.68) at room temperature. The mixture was vigorously vortexed to produce a suspension of multilamellar vesicles (MLVs). To obtain more uniform vesicles, the suspension was first subjected to bath sonication for 10 min and then extruded through 100 nm and 50 nm polycarbonate membranes using a mini-extruder (Avanti Polar Lipids).

\subsection{Sample Preparation}
Liquid crystal droplets were prepared by dispersing monoolein-doped 8CB in aqueous lipid vesicle suspensions. Monoolein was first dissolved in chloroform to prepare a stock solution. A defined amount of the monoolein stock solution was then added to 8CB to obtain liquid crystal samples containing the desired monoolein concentration (1, 2, or 5 wt.-\%). The mixture was vortexed thoroughly to ensure homogeneous mixing of monoolein with 8CB. After mixing, the sample was placed in a vacuum desiccator to remove residual chloroform.

To prepare liquid crystal droplets, a small amount of the monoolein-doped 8CB was added to the lipid vesicle suspension. The mixture was then gently shaken by hand to disperse the liquid crystal phase into the aqueous medium, forming polydisperse liquid crystal droplets. The resulting droplet suspension was subsequently transferred to a glass sample cell and placed under a microscope for observation.

\subsection{Optical Characterization}
The phase transitions and morphological evolution of the liquid crystal droplets were characterized using a research-grade polarized light microscope (AmScope PZ620) operating in transmission mode. To regulate the sample temperature during the heating cycles, the microscope was equipped with a Linkam CO102 temperature-controlled stage. The samples were heated at a rate of 1~$^\circ$C min$^{-1}$. All observations were conducted under crossed polarizers to resolve the liquid crystal director field and birefringent textures.

Images and time-lapse videos were captured using an AmScope MU1803 digital camera (18.0 MP, CMOS sensor) at a resolution of 4912 × 3684 pixels. A 40× infinity-corrected plan-achromatic objective with a numerical aperture (NA) of 0.65 and a working distance (WD) of 0.54 mm was employed for all imaging. Spatial calibration was performed using a stage micrometer, and subsequent image analysis was carried out using the Fiji distribution of ImageJ.

\subsection{Interfacial Tension Measurements}
The liquid crystal-aqueous interfacial tension was measured by the pendant-drop method using a DataPhysics OCA optical contact angle meter (DataPhysics Instruments, Germany) equipped with SCA 20 software. Because 8CB is extremely viscous at room temperature, 5CB was used as a proxy for these measurements. A drop of 5CB was formed at the tip of a needle and immersed in the surrounding aqueous phase, and the equilibrium drop profile was fitted to the Young-Laplace equation to obtain the interfacial tension. Interfacial tensions were measured both in lipid-free buffer (10 mM Tris with 0.1 M NaCl at pH 9.68) and in the same buffer containing 0.5 or 1 mM DLPG or DOPG, as reported in the main text (Fig.~5). All measurements were performed at room temperature. Each value reported in Fig.~5 is the mean of five independent droplets, and the error bars represent the standard deviation.

\begin{acknowledgments}

M.L. acknowledges support from the China Scholarship Council. M.L. acknowledges the use of Anthropic's Claude for assistance with the Python and LaTeX scripts used for figure preparation and typesetting, and for drafting the text. All experimental work, parameters, and reported values were determined and verified by the authors. All authors have reviewed and edited the final work and take full responsibility for the content. L.T. acknowledges support from the European Commission (Horizon-MSCA, Grant No. 892354), the Dutch Research Council NWO ENW Veni grant (Project No. VI.Veni.212.028), and the Dutch Research Council NWO ENW M grant (Project No. OCENW.M.23.360).

\end{acknowledgments}

\section*{Author Contributions}
M.L.: conceptualization, investigation, methodology, formal analysis, visualization, writing---original draft. L.T.: conceptualization, supervision, funding acquisition, writing---review and editing. Both authors have given approval to the final version of the manuscript.

\section*{Notes}
The authors declare no competing financial interest.

\section*{Data Availability}
The time-lapse polarized optical microscopy video dataset (Movies S1-S9) that supports the findings of this study is openly available in DataverseNL at \url{https://doi.org/10.34894/VXJ17C}.

\begin{suppinfo}

\setcounter{figure}{0}
\renewcommand{\thefigure}{S\arabic{figure}}

\noindent The Supporting Information includes polarized optical microscopy images of monoolein-doped 8CB droplets in lipid-free buffer (Fig.~S1) and in low-concentration DLPG solutions (Figs.~S2-S4); droplets in DOPG solutions (Fig.~S5) and in mixed DLPG:DOPG (1:1) solutions (Fig.~S6); chemical structures of DLPG and DOPG (Fig.~S7); and a QR code linking to the supplementary video dataset, Movies S1-S9 (Fig.~S8).

\begin{figure}[htbp]

 \centering
\centering
\includegraphics[width=1\linewidth]{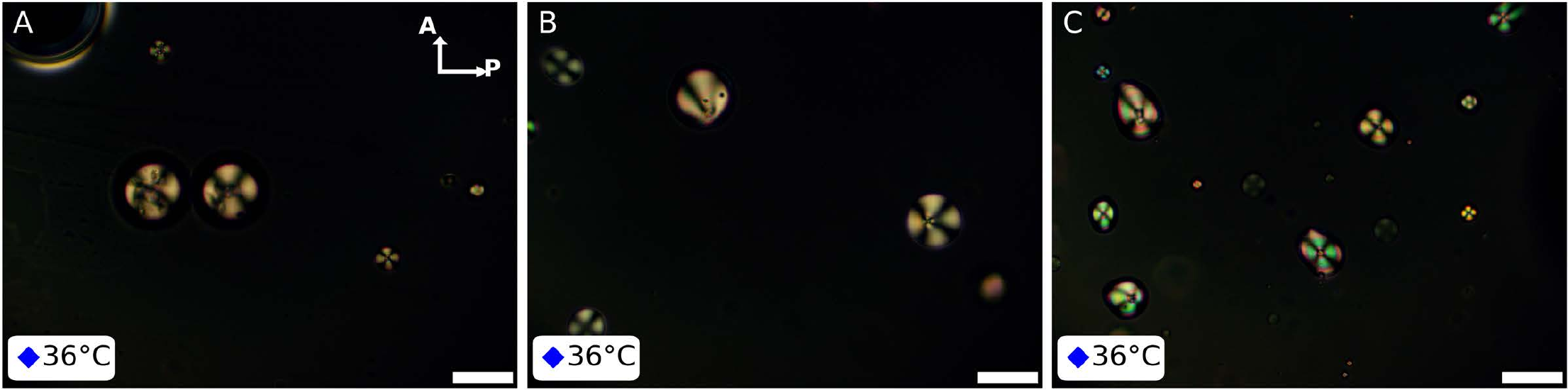}
\caption{Polarized optical microscopy images of monoolein-doped 8CB droplets in DLPG (0mM) buffer solution at 36~$^\circ$C. (A) 1 wt.-\% monoolein, (B) 2 wt.-\% monoolein, and (C) 5 wt.-\% monoolein. All samples exhibit nematic textures under crossed polarizers. Scale bars: 100 μm.}
\label{fig:si-buffer}
\end{figure}

\begin{figure}[htbp]

 \centering
\centering
\includegraphics[width=1\linewidth]{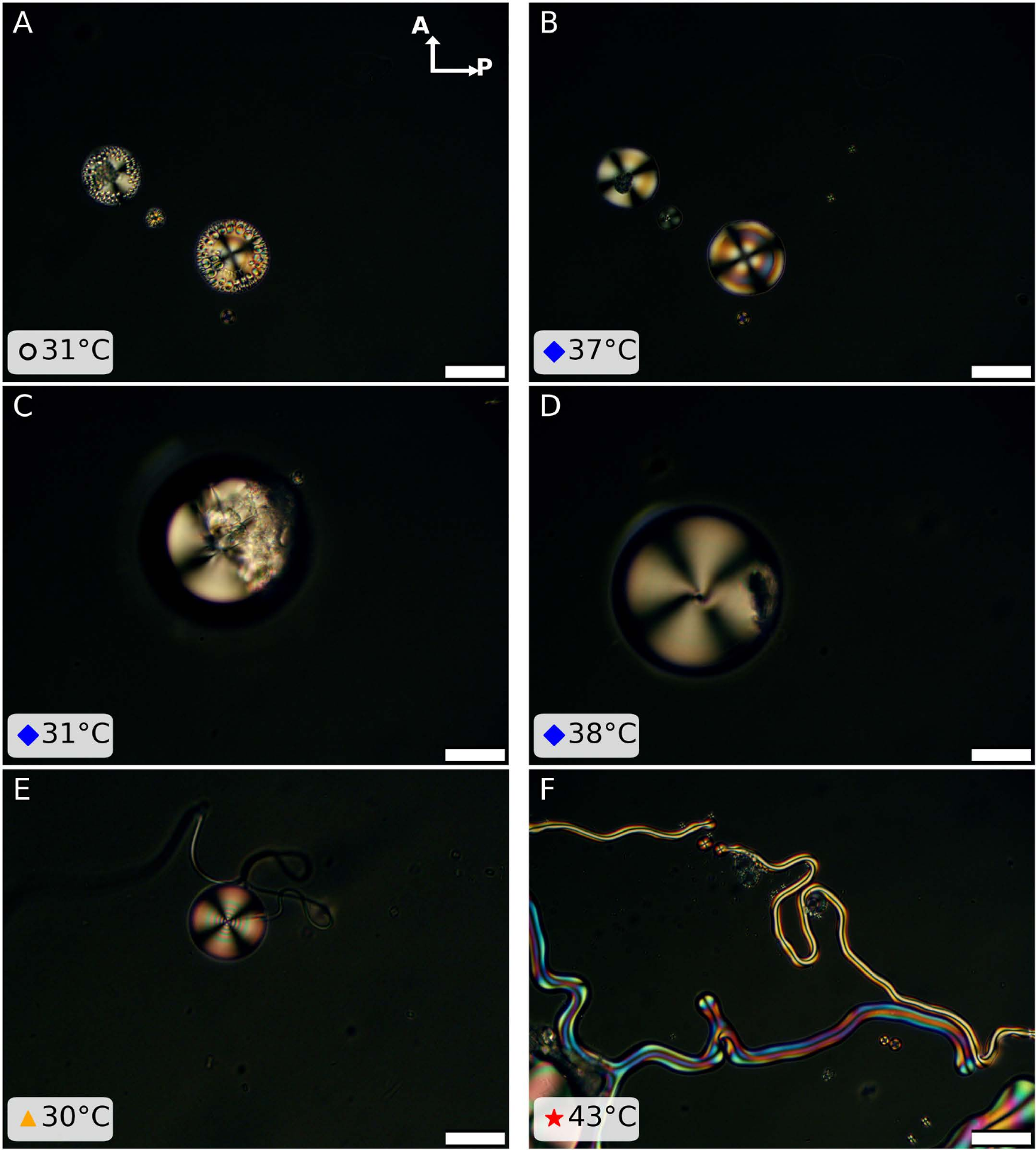}
\caption{Polarized optical microscopy images of 1 wt.-\% monoolein-doped 8CB droplets in aqueous DLPG solutions at low lipid concentrations. (A,B) 0.01 mM DLPG, showing smectic and nematic states at 31 and 37~$^\circ$C, respectively. (C,D) 0.1 mM DLPG, showing nematic textures at 31 and 38~$^\circ$C. (E,F) 0.5 mM DLPG, showing smectic filaments at 30~$^\circ$C and a shape-change state at 43~$^\circ$C. Scale bars: 100 μm.}
\label{fig:si-dlpg1wt}
\end{figure}

\begin{figure}[htbp]

 \centering
\centering
\includegraphics[width=1\linewidth]{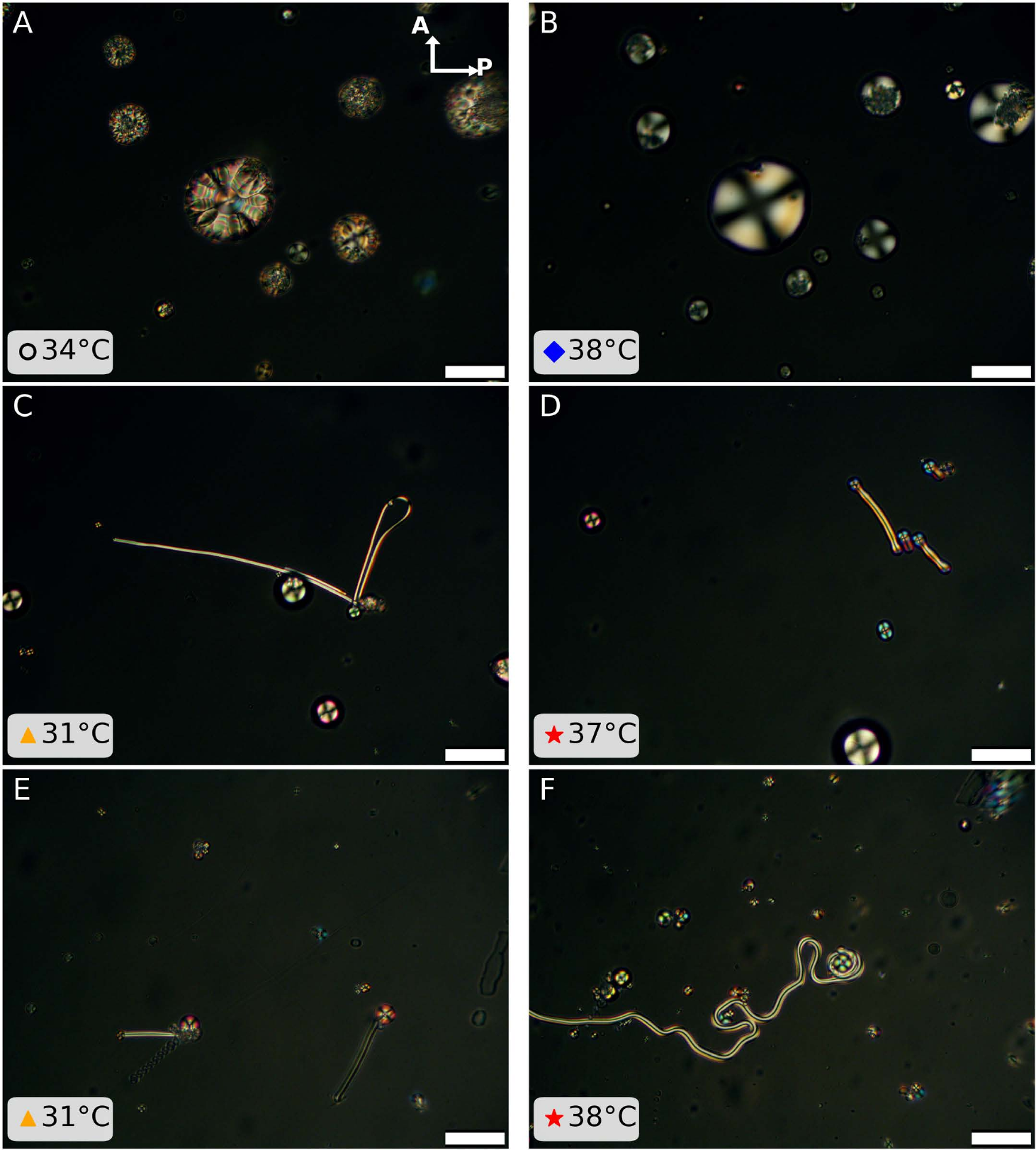}
\caption{Polarized optical microscopy images of 2 wt.-\% monoolein-doped 8CB droplets in aqueous DLPG solutions at low lipid concentrations. (A,B) 0.01 mM DLPG, showing smectic and nematic states at 34 and 38~$^\circ$C, respectively. (C,D) 0.1 mM DLPG, showing smectic filaments at 31~$^\circ$C and a shape-change state at 37~$^\circ$C. (E,F) 0.5 mM DLPG, showing smectic filaments at 31~$^\circ$C and a shape-change state at 38~$^\circ$C. Scale bars: 100 μm.}
\label{fig:si-dlpg2wt}
\end{figure}

\begin{figure}[htbp]

 \centering
\centering
\includegraphics[width=1\linewidth]{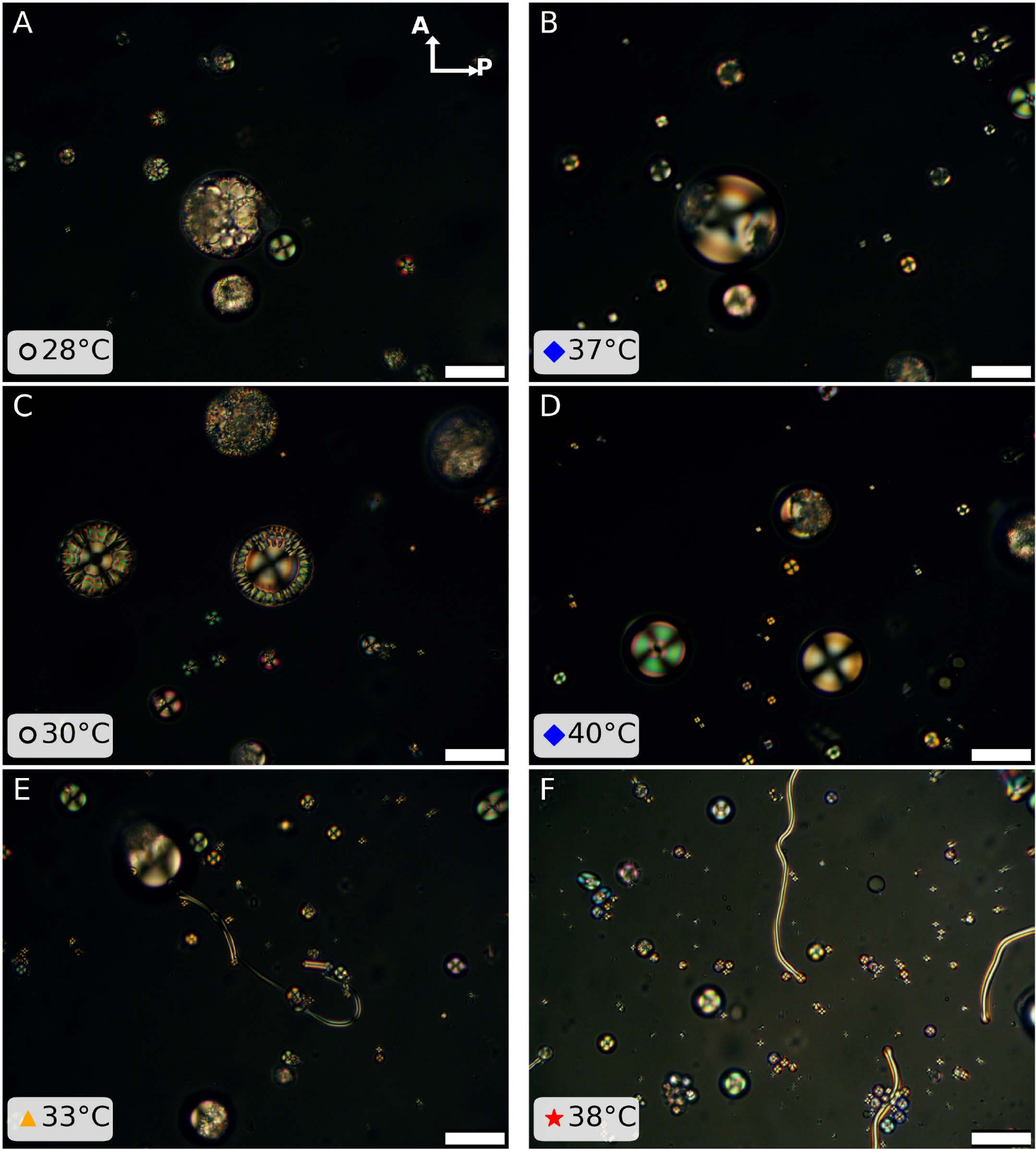}
\caption{Polarized optical microscopy images of 5 wt.-\% monoolein-doped 8CB droplets in aqueous DLPG solutions at low lipid concentrations. (A,B) 0.01 mM DLPG, showing smectic and nematic states at 28 and 37~$^\circ$C, respectively. (C,D) 0.1 mM DLPG, showing smectic and nematic states at 30 and 40~$^\circ$C, respectively. (E,F) 0.5 mM DLPG, showing smectic filaments at 33~$^\circ$C and a shape-change state at 38~$^\circ$C. Scale bars: 100 μm.}
\label{fig:si-dlpg5wt}
\end{figure}

\begin{figure}[htbp]

 \centering
\centering
\includegraphics[width=1\linewidth]{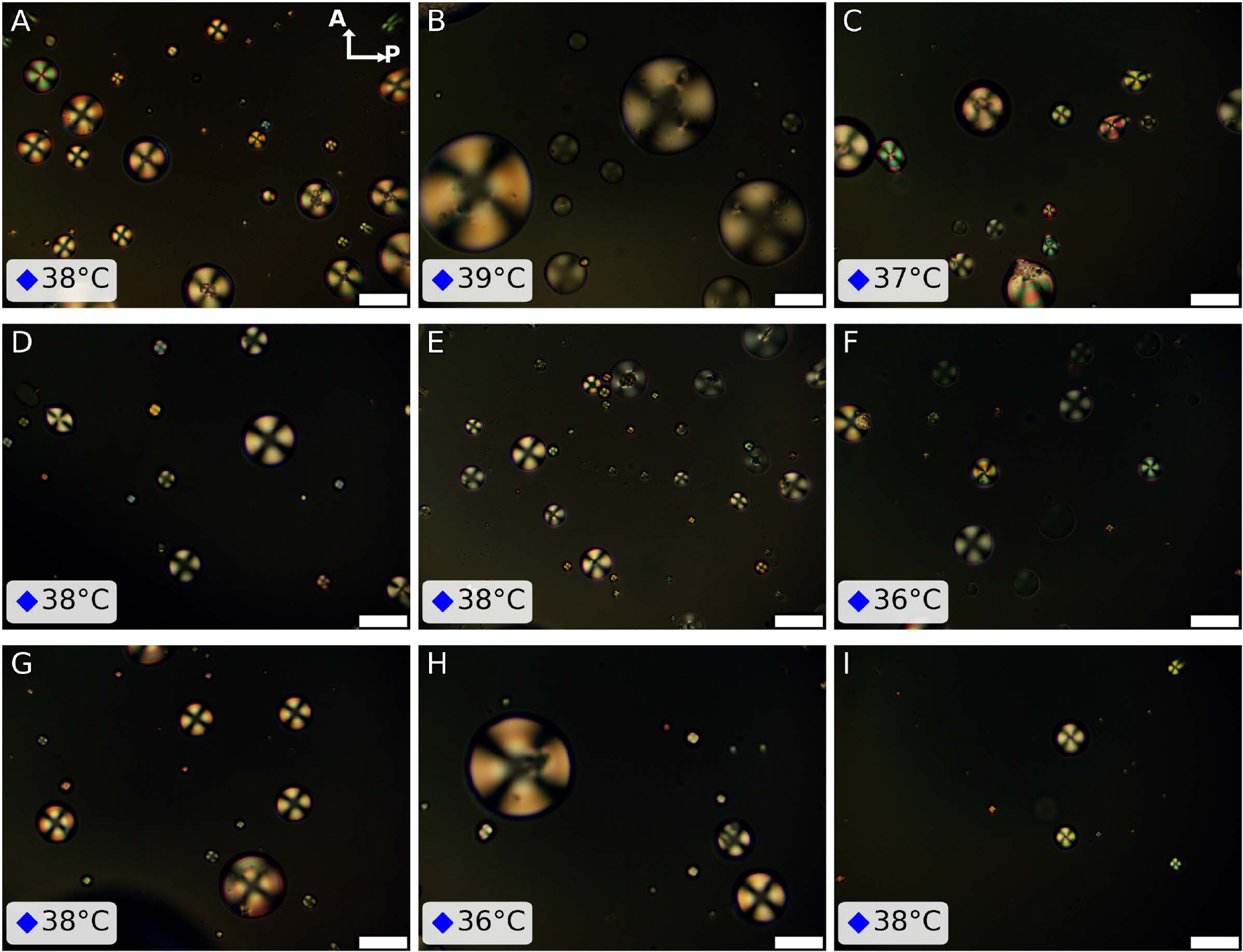}
\caption{Polarized optical microscopy images of monoolein-doped 8CB droplets in aqueous DOPG solutions at different lipid concentrations and monoolein contents. (A-C) Droplets in 1 mM DOPG with 1, 2, and 5 wt.-\% monoolein, respectively. (D-F) Droplets in 5 mM DOPG with 1, 2, and 5 wt.-\% monoolein, respectively. (G-I) Droplets in 10 mM DOPG with 1, 2, and 5 wt.-\% monoolein, respectively. All panels show representative nematic textures observed under crossed polarizers. Scale bars: 100 μm.}
\label{fig:si-dopg}
\end{figure}

8CB droplets in aqueous DOPG are shown in Fig.~S5. At 1 mM DOPG, the droplets exhibited nematic textures at 38~$^\circ$C for 1 wt.-\% monoolein, 39~$^\circ$C for 2 wt.-\% monoolein, and 37~$^\circ$C for 5 wt.-\% monoolein. At 5 mM DOPG, nematic droplets were observed at 38~$^\circ$C for both 1 wt.-\% and 2 wt.-\% monoolein and at 36~$^\circ$C for 5 wt.-\% monoolein. At 10 mM DOPG, the droplets also showed nematic textures, observed at 38~$^\circ$C for 1 wt.-\% monoolein, 36~$^\circ$C for 2 wt.-\% monoolein, and 38~$^\circ$C for 5 wt.-\% monoolein. In all these cases, the droplets remained spherical and displayed typical nematic birefringent textures under crossed polarizers, without the smectic filaments or shape-change behavior observed in the corresponding DLPG systems.

\begin{figure}[htbp]

 \centering
\centering
\includegraphics[width=1\linewidth]{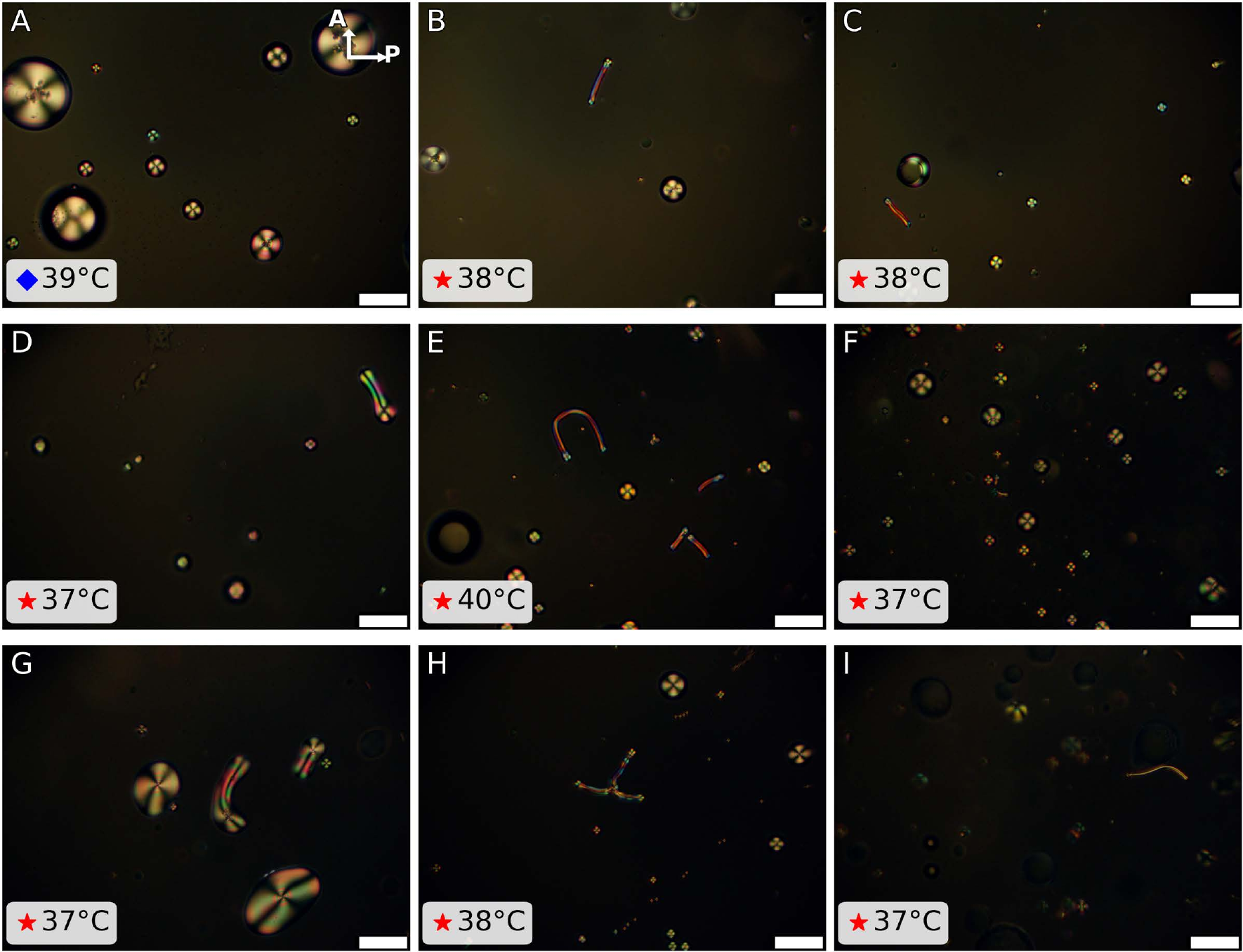}
\caption{Polarized optical microscopy images of monoolein-doped 8CB droplets in aqueous mixed lipid solutions with a DLPG:DOPG ratio of 1:1 at different lipid concentrations and monoolein contents. (A-C) Droplets in 1 mM mixed lipid solution with 1, 2, and 5 wt.-\% monoolein, respectively. (D-F) Droplets in 5 mM mixed lipid solution with 1, 2, and 5 wt.-\% monoolein, respectively. (G-I) Droplets in 10 mM mixed lipid solution with 1, 2, and 5 wt.-\% monoolein, respectively. Scale bars: 100 μm.}
\label{fig:si-mix}
\end{figure}

Micrographs of monoolein-doped 8CB droplets in mixed DLPG:DOPG (1:1) solutions are shown in Fig.~S6. At 1 mM mixed lipid, a nematic texture was observed for 1 wt.-\% monoolein at 39~$^\circ$C, whereas shape-change structures were observed for both 2 wt.-\% and 5 wt.-\% monoolein at 38~$^\circ$C. At 5 mM mixed lipid, shape-change behavior was observed at 37~$^\circ$C for 1 wt.-\% monoolein, at 40~$^\circ$C for 2 wt.-\% monoolein, and at 37~$^\circ$C for 5 wt.-\% monoolein. At 10 mM mixed lipid, shape-change structures were observed at 37.5~$^\circ$C for 1 wt.-\% monoolein, at 38~$^\circ$C for 2 wt.-\% monoolein, and at 37~$^\circ$C for 5 wt.-\% monoolein. Compared with the DOPG-only system, the mixed lipid system showed more evident shape-change behavior under these conditions. However, compared with the DLPG-only system, the mixed lipid system showed less evident phase-transition behavior under similar conditions.

\begin{figure}[htbp]

 \centering
\centering
\includegraphics[width=0.7\linewidth]{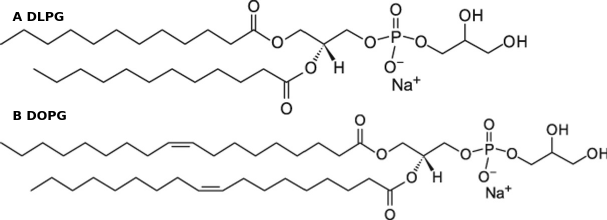}
\caption{Chemical structures of the phospholipids used in this study. (A) 1,2-dilauroyl-\textit{sn}-glycero-3-phosphoglycerol (DLPG). (B) 1,2-dioleoyl-\textit{sn}-glycero-3-phosphoglycerol (DOPG).}
\end{figure}

\begin{figure}[htbp]

 \centering
\centering
\includegraphics[width=0.5\linewidth]{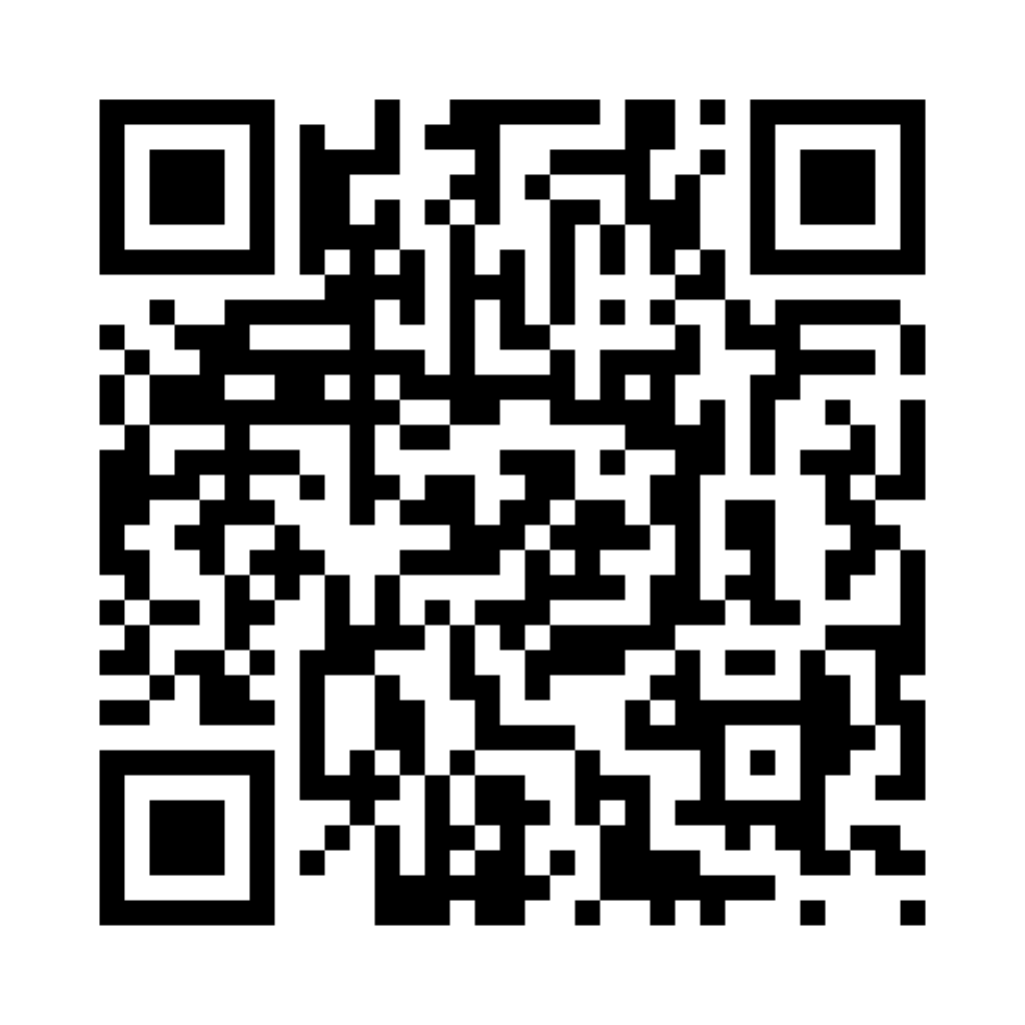}
\caption{QR code linking to the supplementary video dataset for this study. Scanning the code opens the dataset
``Lipid-Mediated Control of Thermally Induced Shape Transformations in Liquid Crystal Droplets'' deposited on Data-
verseNL (DOI: 10.34894/VXJ17C). The dataset contains Movies S1–S9, time-lapse polarized optical micrographs of
monoolein-doped 8CB liquid-crystal droplets in aqueous DLPG solutions recorded during temperature scans.}
\end{figure}
\end{suppinfo}

\clearpage
\bibliography{references}

\end{document}